\documentclass[showpacs,twocolumn,superscriptaddress]{revtex4}

\usepackage{doi}%<----------
\usepackage{hyperref}
\hypersetup{
  colorlinks=true,        % false: boxed links; true: colored links
  linkcolor=blue,         % color of internal links
  citecolor=cyan,         % color of links to bibliography
}

\usepackage{graphicx}% Include figure files
\usepackage{dcolumn}% Align table columns on decimal point
\usepackage{bm}% bold math
\usepackage{color}

\usepackage{amsmath}
\usepackage{amssymb}

\begin{document}

\title{Charged particle motion around a magnetized Reissner-Nordstr\"{o}m black hole }

\author{Sanjar Shaymatov}
\email{sanjar@astrin.uz}

\affiliation{Ulugh Beg Astronomical Institute, Astronomy St. 33, Tashkent 100052, Uzbekistan}
\affiliation{Akfa University, Kichik Halqa Yuli Street 17, Tashkent 100095, Uzbekistan}
\affiliation{Institute for Theoretical Physics and Cosmology, Zheijiang University of Technology, Hangzhou 310023, China}
\affiliation{National University of Uzbekistan, Tashkent 100174, Uzbekistan}
\affiliation{Tashkent Institute of Irrigation and Agricultural Mechanization Engineers, Kori Niyoziy 39, Tashkent 100000, Uzbekistan}

\author{Bakhtiyor Narzilloev}
\email{nbakhtiyor18@fudan.edu.cn}
\affiliation{Center for Field Theory and Particle Physics and Department of Physics, Fudan University, 200438 Shanghai, China }
\affiliation{Akfa University, Kichik Halqa Yuli Street 17, Tashkent 100095, Uzbekistan}
\affiliation{Ulugh Beg Astronomical Institute, Astronomy St. 33, Tashkent 100052, Uzbekistan}

\author{Ahmadjon~Abdujabbarov}
\email{ahmadjon@astrin.uz}
\affiliation{Shanghai Astronomical Observatory, 80 Nandan Road, Shanghai 200030, China}
\affiliation{Ulugh Beg Astronomical Institute, Astronomy St. 33, Tashkent 100052, Uzbekistan}
\affiliation{Institute of Nuclear Physics, Ulugbek 1, Tashkent 100214, Uzbekistan}
\affiliation{National University of Uzbekistan, Tashkent 100174, Uzbekistan}
\affiliation{Tashkent Institute of Irrigation and Agricultural Mechanization Engineers, Kori Niyoziy 39, Tashkent 100000, Uzbekistan}

\author{Cosimo Bambi}
\email{bambi@fudan.edu.cn}
\affiliation{Center for Field Theory and Particle Physics and Department of Physics, Fudan University, 200438 Shanghai, China }

\date{\today}
\begin{abstract}
We investigate the dynamics of neutral and charged test particles around axially symmetric magnetized black hole spacetime. We consider its electromagnetic field in the black hole vicinity and study its impact on the dynamics of test particles. We determine the radius of the innermost stable circular orbit (ISCO) for neutral and charged test particles and show that the combined effect of black hole electric charge and magnetic field strongly affects the ISCO radius, thus shrinking its values. We also show that the ISCO radius of positively (negatively) charged particle initially gets increased (decreased) and then gets radically altered with an increase in the value of both black hole electric charge and test particle charge. It turns out that the repulsive (attractive) Coulomb force dominates over the Lorentz force arising from the black hole magnetic field. 
Typically, black hole rotation causes axially symmetric spacetime case. Similarly, it turns out that a magnetized black hole solution also causes axially symmetric spacetime as a consequence of the presence of magnetic field. We study a degeneracy for the value of the ISCO between the Kerr and the magnetized Reissner-Nordstr\"{o}m black hole geometries and show that the combined effects of black hole charge and magnetic field can be mimicked by Kerr spacetime with the spin parameter up to $a/M\approx 0.8$.  Finally, we consider the center of mass energy of colliding particles and show that an increase in the values of black hole magnetic field and electric charge leads to high center of mass energy extracted by collision of two particles.

\end{abstract}
\maketitle

\section{Introduction}
\label{introduction}

Typically in an astrophysical scenario black holes are believed to be surrounded by a magnetic field. In this sense, most of the previous analyses were devoted to investigate the motion of charged particles in the model of central gravitational object 
placed in an external magnetic field being weak that does not modify the background black hole geometry. 
Thus, it is regarded as a test field for analytical and numerical analyses in the given gravitational background~
\cite[see, e.g.][]{Frolov10,Aliev02,Abdujabbarov10,Shaymatov14,Shaymatov18a,Kolos15,Stuchlik16,Tursunov16}.
The strength of the magnetic field $B$ was estimated for both stellar mass black holes and supermassive black holes \cite{Piotrovich10},
and its strength at the horizon radius of the black hole
was also measured~\cite{Eatough13,Shannon13}. 
The point to be noted here is that the gravitational energy is comparable with the electromagnetic energy in the case when the strength of magnetic field surrounding black hole with mass $M$ is of order $ B\sim B_{max} \sim 10^{19}\frac{M_\odot}{M}~ \rm{Gauss}$. \textcolor{black}{It is worth noting that $B_{max}$ is referred to as the upper limit of magnetic field that makes black hole \textit{strongly magnetized}.  However, the magnetic field strength  is of order $B_1\sim 10^{8}~\rm{Gauss}$ for stellar mass black holes and $B_2\sim 10^{4}~\rm{Gauss}$ for the supermassive black holes~\cite{Piotrovich10}. Later on, it was estimated to be between $200$ and $8.3 \times 10^{4}$~\rm{Gauss} at 1 Schwarzschild radius~\cite{Baczko16}. Also, recent observational analysis of binary black  hole  system $V404$  Cygni suggests that the magnetic field is of order $B\sim 33.1 \pm 0.9~\rm{Gauss}$ in the
corona~\cite{Dallilar2018}. Thus, the magnetic field
in the black hole vicinity is considered to be much smaller, i.e. $B_{1,2}\ll B_{max}$. But the magnetic field can affect the motion of charged
particles drastically as that of the large Lorentz force, regardless of its small value.}  However, investigations suggest that the spacetime geometry near a black hole would be significantly distorted provided that the strength of the magnetic field is of order $B_{max}$. For that reason, a strong magnetic field would be very important as a background field testing the black hole geometry. Recently, properties of the magnetized Reissner-Nordstr\"{o}m black hole solution was investigated in Ref.~\cite{Gibbons13}.

Particularly, the family of solution describing so-called magnetized black holes has been suggested by authors of Ref.~\cite{Aliev89}. These solutions involve the interaction of the gravity with the axially symmetric magnetic field produced by an external source, e.g., by the accretion disk surrounding black hole. Another interesting solution describing the magnetized black hole has been introduced in Ref.~\cite{Ernst76}, considering a nonlinear coupling the Schwarzschild black hole with the Melvin’s magnetic universe. This solution represents additional gravity due to the self-gravitating strong asymptotically uniform magnetic field. Other generalizations of such configurations with some complicated asymptotic behaviour have been proposed by the authors of Refs.~\cite{Ernst76wz, Aliev89wz,Garcia85wz}, including rotating and charged magnetized black hole solutions. Later, the authors of Refs.~\cite{Gibbons14wz, Astorino16wz} also attempted to involve the global charge in obtaining the magnetized black hole solutions. 
Overall these above mentioned and other attempts to consider and explore the magnetized black holes are motivated by the several astrophysical processes. One of them is the accretion disk by which the magnetic field can be produced. That would therefore play an important role for formation of the relativistic jets from active galactic nuclei and microquasars. The most relevant model proposed by Blandford and Znajek~\cite{Blandford1977} assumes that the magnetic field is dragged by the rotating black hole and pushes out the plasma along the axis of rotation that could form the relativistic jets.   

On the other hand, one needs to test the solution of general relativity or modified/alternative theories of gravity on the basis of some astrophysical processes. As an example, x-ray data from astrophysical compact objects maybe used as a useful test of the particular theory and corresponding solution~\cite{Bambi12a,Bambi16b,Zhou18,Tripathi19}. One may also use a test particle motion around the black hole in modelling some processes associated with the accretion disc~\cite{Bambi17e,Chandrasekhar98}. Electromagnetic field surrounding the black hole therefore plays the crucial determining role in analyzing the motion of charged particles and other relevant astrophysical processes in the black hole vicinity. The effects of electromagnetic field and modifications done to the gravity on some astrophysical processes have been widely explored in Refs.~ \cite{Chen16,Hashimoto17,Dalui19,Han08,Moura00,Morozova14}. A big number of research works has been devoted to the study of the effect of electromagnetic field and parameters of the compact objects on the dynamics of the charged particles (see for example~\cite{Stuchlik20,Jawad16,Hussain15,Jamil15,Hussain17,Babar16,Banados09,Majeed17,Zakria15,Brevik19,
DeLaurentis2018PhRvD,Shaymatov13,Atamurotov13a,Kolos17,Kovar10,
Kovar14,Aliev86,Frolov11,Frolov12,Stuchlik14a,Abdujabbarov11a,
Abdujabbarov11,Abdujabbarov08,Karas12a,Shaymatov15,Duztas-Jamil20,
Turimov18b,Turimov17,Shaymatov20egb,Shaymatov19b,Shaymatov20b,Narzilloev19,Narzilloev:2020b,Narzilloev20c,Narzilloev20a,
Narzilloev20b,Narzilloev21,Shaymatov21b}). On the other hand, the investigation of the effects of electromagnetic field on particles having a nonzero magnetic dipole in the gravitational field of compact object has been explored in Refs.~\cite{deFelice,defelice2004,Toshmatov15d,Abdujabbarov14,Rahimov11a,Rahimov11,
Haydarov20,Rayimbaev20c,Vrba20,Abdujabbarov2020jla,TurimovPhysRevD2020,
BokhariPhysRevD2020,Shaymatov20d}.

Here, we plan to explore the charged particle dynamics in the magnetized Reissner-Nordstr\"om black hole vicinity. In Sec.~\ref{Sec:Magnetized} we briefly describe the magnetized black hole spacetime and properties of its electromagnetic field. In
Sec.~\ref{Sec:Motion} we study the motion of charged particles in the strong gravitational and electromagnetic fields of the magnetized black hole. In Sec.~\ref{Sec:4} we focus on the collision of test particles around magnetized Reissner-Nordstr\"{o}m black hole. Sec.~\ref{Sec:conclusion} is devoted to the concluding remarks of
the obtained results. Throughout the manuscript we use a system of units in which $G=c=1$.

\section{Magnetized Reissner-Nordstr\"{o}m black hole metric and its electromagnetic field }\label{Sec:Magnetized}

The spacetime metric of the magnetized Reissner-Nordstr\"{o}m black hole in
the Boyer-Lindquist coordinates is given by \cite{Gibbons13}
\begin{eqnarray}\label{Eq:metric} d s^2 &=& H\, [-f dt^2 + f^{-1}\, dr^2  + r^2 d\theta^2] +
      H^{-1}\, r^2\sin^2\theta\, \nonumber\\
 &&\times(d\phi -\omega dt)^2\, ,
  \end{eqnarray}
%%%%%
where \begin{eqnarray}\label{Eq:fhw}
f&=& 1- \frac{2M}{r} + \frac{Q^2}{r^2}\, ,\nonumber\\
H &=& 1 +\frac{1}{2}B^2 (r^2\sin^2\theta + 3 Q^2\cos^2\theta)
\nonumber\\&&+
  \frac{1}{16} B^4 (r^2 \sin^2\theta + Q^2\cos^2\theta)^2\, ,\nonumber\\
\omega &=& -\frac{2Q B}{r} + \frac{1}{2} Q B^3\, r (1+f
\cos^2\theta)\, , \end{eqnarray}
\textcolor{black}{where $M$ and $Q$ are black hole mass and electric charge respectively while $B$ is the parameter describing the magnetic field. } 

The event horizon of the magnetized black hole is defined as 
 \begin{eqnarray}
 r_{h}=M^2+\sqrt{{M}^2-Q^{2}}\,.
 \end{eqnarray}
It is worth noting that the radius of the event horizon of the magnetized
Reissner-Nordstr\"{o}m black hole remains the same as compared to the Reissner-Nordstr\"{o}m black hole case. 

Vector potential of the electromagnetic field around the black hole has the form as
\begin{eqnarray}\label{Eq:4-vec}
A &=& A_t dt + A_\phi (d\phi-\omega dt)\, , 
\end{eqnarray}
where components of the vector potential of the
electromagnetic field are given as follows: 
 \begin{eqnarray}
A_t &=& -\frac{Q}{r} +\frac{3}{4} Q B^2 r\, (1+ f\cos^2\theta)\,,\nonumber\\
A_\phi &=& \frac{2}{B} - H^{-1}\Big[\frac{2}{B} +
             \frac{1}{2} B(r^2\sin^2\theta + 3 Q^2
             \cos^2\theta)\Big]\,. \nonumber\\
\end{eqnarray}
%%%%%

Since there is a nondiagonal component in the magnetized
Reissner-Nordstr\"{o}m spacetime metric we consider zero angular momentum observer (ZAMO). For ZAMO, the four-velocity components take the following form
\begin{eqnarray}
\label{zamo_con} &(u_{_{\textrm{ZAMO}}})^{\alpha}
=\left\{(Hf)^{-1/2},0,0,\omega(Hf)^{-1/2}\right\}\
, \\
\label{zamo_cov} &(u_{_{\textrm{ZAMO}}})_{\alpha}=
\left\{(Hf)^{1/2},0,0,0\right\}\, .
\end{eqnarray}
The nonvanishing components of the Faraday tensor are given as
follows
%
%\begin{widetext}
\begin{eqnarray}
\label{f1} F_{rt}&=&-\frac{Q}{r^2}+\frac{3}{4}QB^2\left( 1+\left(1-\frac{Q^2}{r^2}\right)\cos^2\theta\right)\, ,\\
\label{f2} F_{\theta t}&=&-\frac{3}{2}QB^2r\left(1-\frac{2M}{r}+\frac{Q^2}{r^2}\right)\sin\theta\cos\theta\, ,  \\
\label{f3}
F_{r\phi}&=&H^{-1}Br\bigg\{H^{-1}\left[1+\frac{B^2}{4}\left(r^2\sin^2\theta
+
Q^2\cos^2\theta\right)\right]\bigg.\nonumber\\
&&\bigg.\times
\left[2+\frac{B^2}{2}\left(r^2\sin^2\theta
+ 3Q^2\cos^2\theta\right)\right]-1\bigg\}\sin^2\theta\, , \\
\label{f4} 
F_{\theta
\phi}&=&H^{-1}B\bigg(H^{-1}\left[r^2-3Q^2+\frac{B^2}{4}\left(
r^2-Q^2\right)\left(r^2\sin^2\theta \bigg.\right.\right. \nonumber\\
&&\bigg.\left.\left.
+ Q^2\cos^2\theta\right)\right]
\left[2+\frac{B^2}{2}\left(r^2\sin^2\theta +
3Q^2\cos^2\theta\right)\right]
\bigg.\nonumber\\
&&\bigg.-r^2+3Q^2\bigg) \sin\theta\cos\theta\,.
\end{eqnarray}
%\end{widetext}
%
From Eqs. %Using the four-velocity components of ZAMO
(\ref{zamo_con}) and (\ref{f4}) %and on the basis of the expressions (\ref{f1}--\ref{f4}) 
one can then write the orthonormal components of the electromagnetic field measured by ZAMO as 
\begin{widetext}
\begin{eqnarray}
\label{e1}E^{\hat r} &&= H^{-1}B\Bigg\{
\left(\frac{Q}{Br^2}+\frac{3}{4}Q
B\left(1+\left(1-\frac{Q^2}{r^2}\right)\cos^2\theta\right)\right)
 +H^{-1}\omega r\left(H^{-1}\left[1+\frac{B^2}{4}\left(r^2\sin^2\theta
+
Q^2\cos^2\theta\right)\right]\right.\Bigg.\nonumber\\
&&\Bigg.\left.\times \left[2+\frac{B^2}{2}\left(r^2\sin^2\theta +
3Q^2\cos^2\theta\right)\right]-1\right)\sin^2\theta\Bigg\} \, ,\\
\label{e2}  E^{\hat \theta}&&
=\frac{BH^{-1}}{r}\Bigg\{f^{-1/2}\omega
H^{-1}\Bigg(H^{-1}\left[r^2-3Q^2+\frac{B^2}{4}\left(
r^2-Q^2\right)\left(r^2\sin^2\theta +
Q^2\cos^2\theta\right)\right]\Bigg.\Bigg.\nonumber\\
&&\Bigg.\Bigg.\times \left[2+\frac{B^2}{2}\left(r^2\sin^2\theta
+
3Q^2\cos^2\theta\right)\right]-r^2+3Q^2\Bigg) -\frac{3}{2}Q B
r\left(\frac{1}{f}\right)^{-1/2}\Bigg\} \sin\theta\cos\theta
 \, ,\\
\label{b1}  B^{\hat r} &&=-\frac{B(f/H)^{1/2}}{r^2\left(H
f-H^{-1}\omega^2r^2\sin^2\theta\right)^{1/2}}\bigg(H^{-1}\left[r^2-3Q^2+\frac{B^2}{4}\left(
r^2-Q^2\right)\left(r^2\sin^2\theta +
Q^2\cos^2\theta\right)\right]\Bigg.\nonumber\\
&&\Bigg.\times\left[2+\frac{B^2}{2}\left(r^2\sin^2\theta +
3Q^2\cos^2\theta\right)\right]
-r^2+3Q^2\bigg)\cos\theta \, , \\
\label{b2}  B^{\hat\theta} && =\frac{B f}{\left(H^2
f-\omega^2r^2\sin^2\theta\right)^{1/2}}\bigg\{H^{-1}\left[1+\frac{B^2}{4}\left(r^2\sin^2\theta
+
Q^2\cos^2\theta\right)\right]\left[2+\frac{B^2}{2}\left(r^2\sin^2\theta
+ 3Q^2\cos^2\theta\right)\right]-1\bigg\}\sin\theta \, .\nonumber\\
\end{eqnarray}
\end{widetext}
\begin{figure*}
 \includegraphics[width=0.25\textwidth]{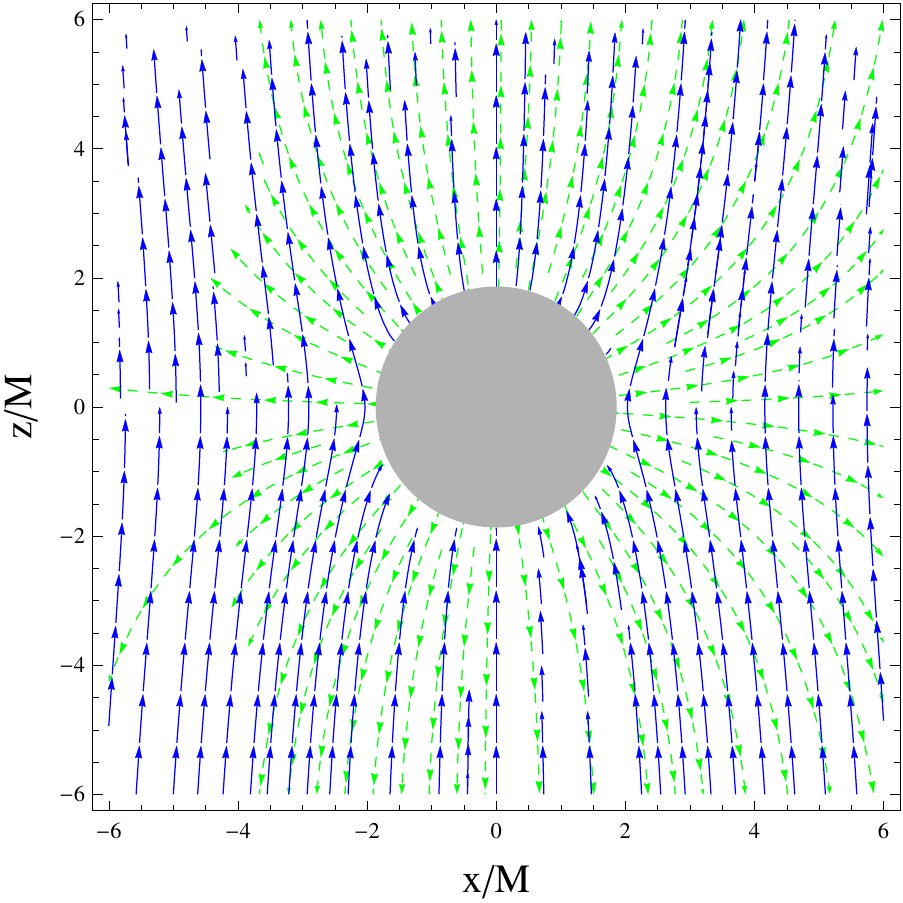}
 \includegraphics[width=0.25\textwidth]{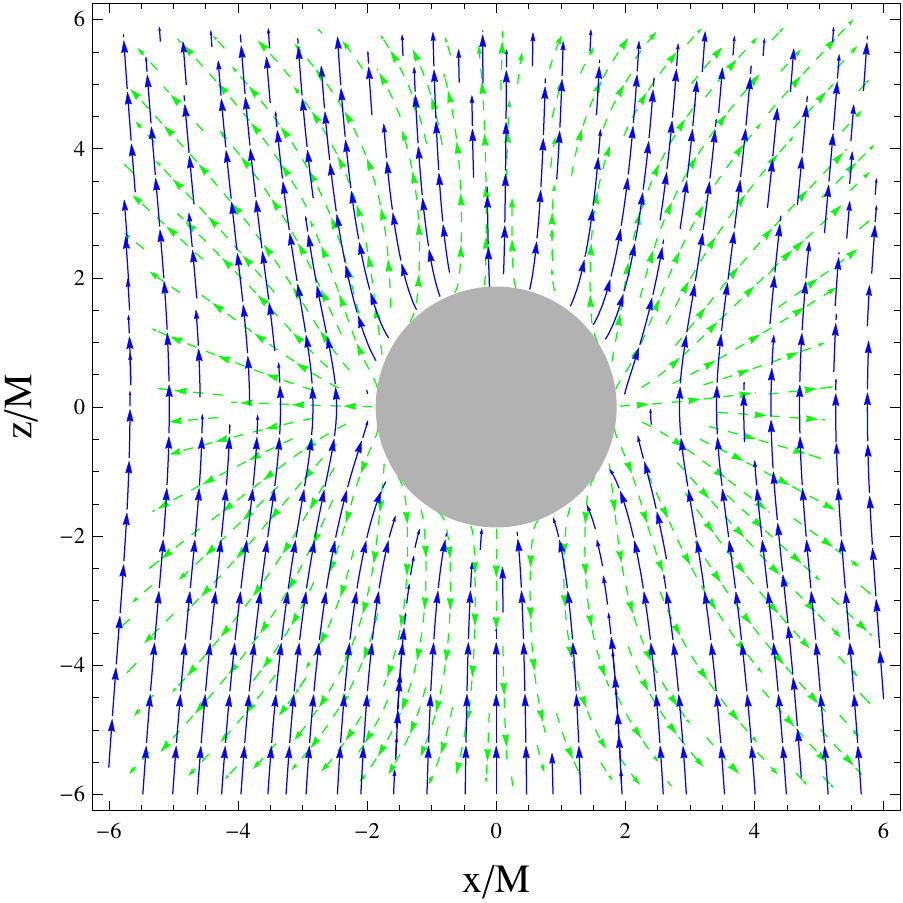}
 \includegraphics[width=0.25\textwidth]{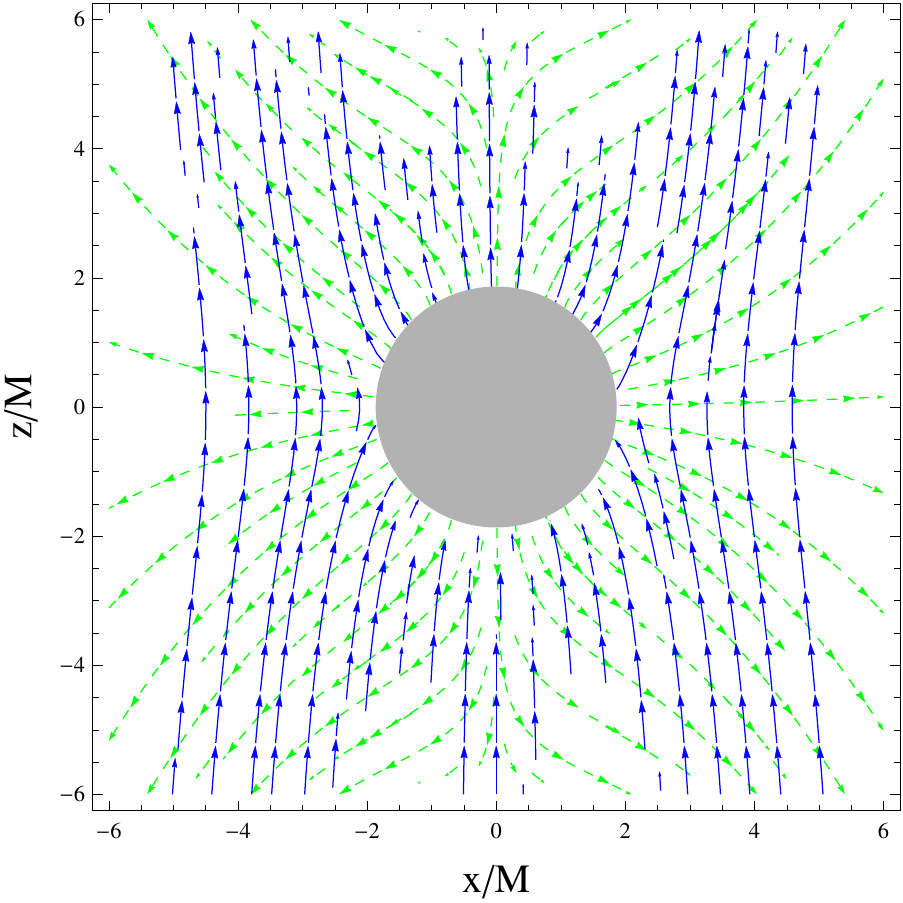} 

 \includegraphics[width=0.25\textwidth]{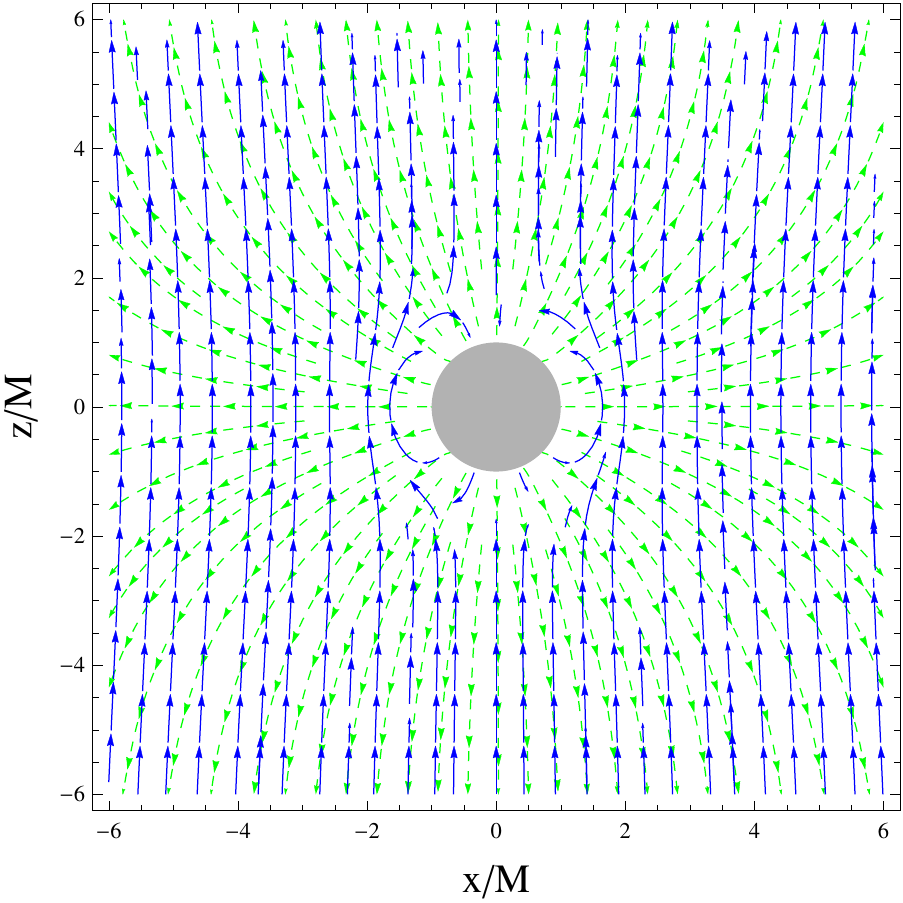}
 \includegraphics[width=0.25\textwidth]{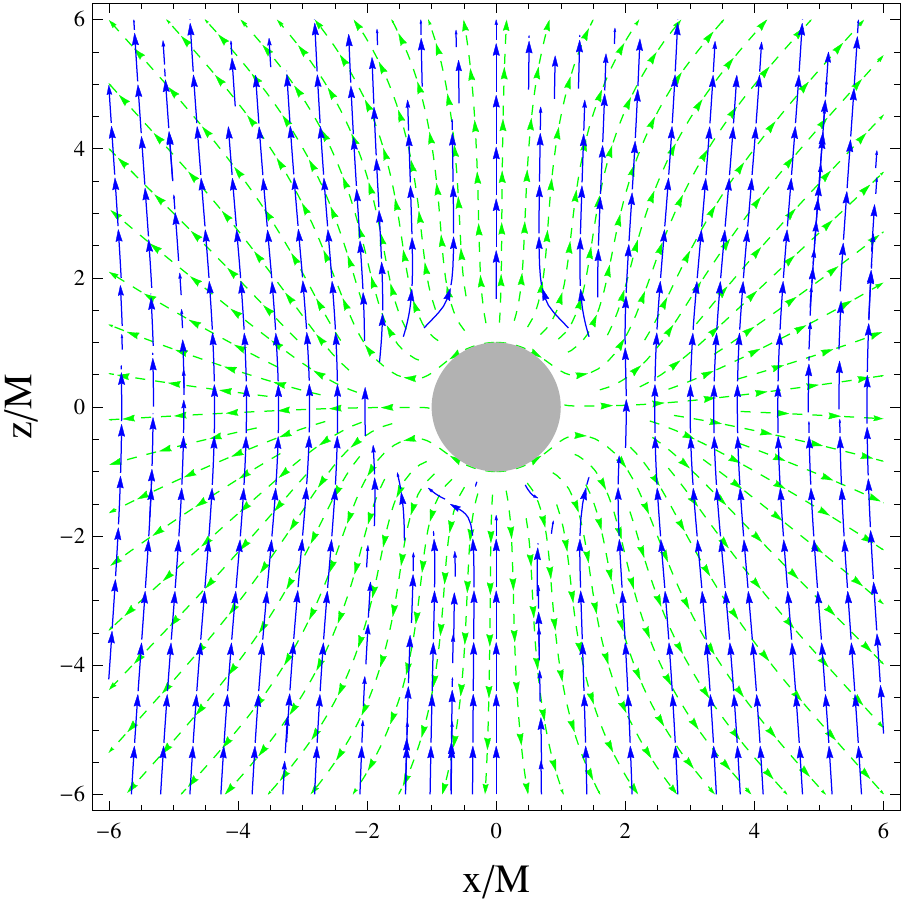}
 \includegraphics[width=0.25\textwidth]{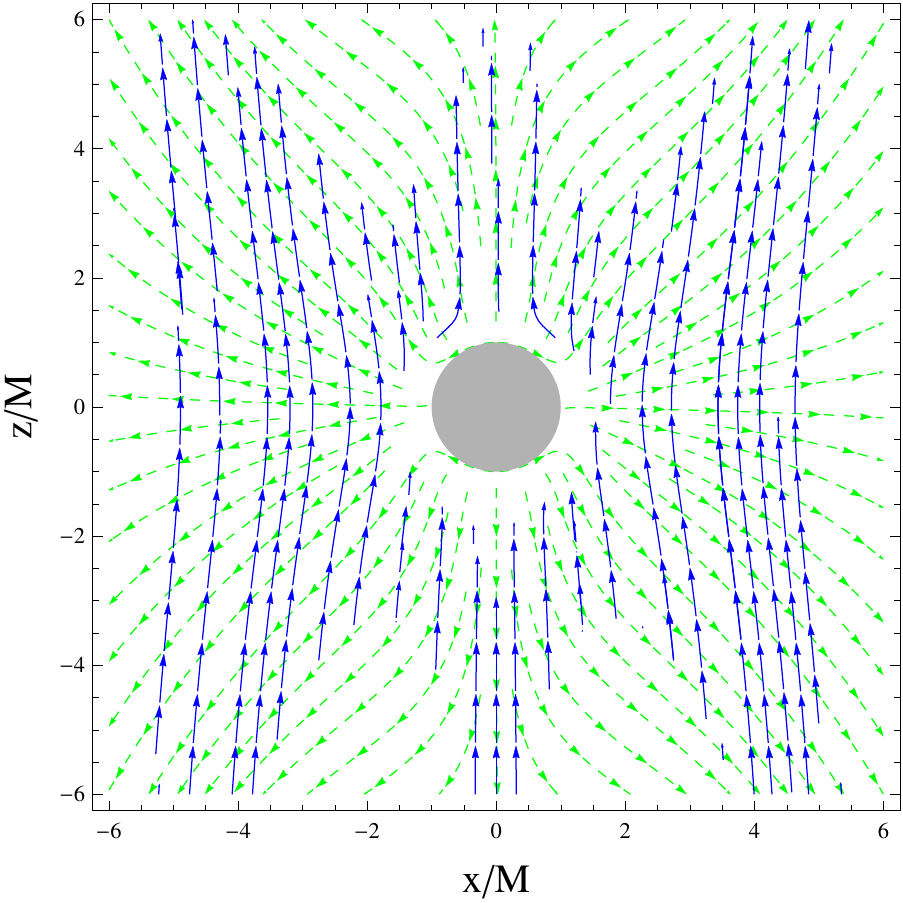} %
\caption{\label{mf1} The configuration of electromagnetic field in the vicinity of a magnetized Reissner-Nordstr\"{o}m black hole. Blue and green lines describe the magnetic and electric fields, respectively. Meanwhile, the horizon is shown as a gray-shaded area. For the figure: first, second, and third columns respectively refer to $B=0.01$, $B=0.05$, and $B=0.1$ in the case of fixed $Q=0.1$ (top panel) and $Q=0.5$ (bottom panel).}
\end{figure*}
In fact, the radial and polar components of the electric field can
appear due to the dragging of inertial frames with dependence from
angular momentum of the black hole. However, in the magnetized
Reissner-Nordstr\"{o}m black hole case, components of the electric field
consist of not only the monopole part associated with the electric charge but also the part of dragging
of inertial frames due to the effect of magnetic filed. In the
limit of flat spacetime, i.e., for \hbox{$M/r\rightarrow 0$,
$Q^2/r^2\rightarrow 0$}, and \hbox{$H\rightarrow 1$}, Eqs.~
(\ref{e1}) and (\ref{b2}) are given by
\begin{eqnarray}
\label{limit_b} && \lim_{M/r,~ Q^2/r^2\rightarrow 0,~ H\rightarrow
1} B^{\hat r} =-B\cos\theta \,, \nonumber
\\ \nonumber \\
 && \lim_{M/r,~ Q^2/r^2\rightarrow 0,~
H\rightarrow 1} B^{\hat\theta}=B\sin\theta   \,, \nonumber
\\ \nonumber \\
\label{limit_e} &&  \qquad \lim_{M/r, ~Q^2/r^2\rightarrow 0, ~
H\rightarrow 1} E^{\hat r,\hat \theta}=0\, .
\end{eqnarray}
As was expected, the above equation coincides with the
solutions for homogeneous magnetic field in Minkowski
spacetime.

The structure of electromagnetic field in the 
magnetized Reissner-Nordstr\"{o}m black hole vicinity is shown in
Fig.~\ref{mf1}. The green and blue vector
field lines respectively refer to the electric and magnetic fields, while the horizon of the black hole is shown as
a gray shaded area. As can be seen from Fig.~\ref{mf1}, the
electromagnetic field lines change as we increase the value of the magnetic field parameter $B$ for given value of black hole charge $Q$. In the next section, we explore a charged particle motion around
the magnetized Reissner-Nordstr\"{o}m black hole.

\section{Particle motion around magnetized Reissner-Nordstr\"{o}m black hole } \label{Sec:Motion} 

In this section, we study a charged particle motion around magnetized Reissner-Nordstr\"{o}m black hole for various cases. 

\textcolor{black}{The Hamiltonian for the charged particle motion is given by \cite{Misner73}
\begin{eqnarray}
H=\frac{1}{2}g^{\mu\nu} \left(\pi_{\mu}-qA_{\mu}\right)\left(\pi_{\nu}-qA_{\nu}\right)\, ,
\end{eqnarray}
where $\pi_{\mu}$ is the canonical momentum of the charged particle and $A_{\mu}$ is the four-vector potential of the electromagnetic field. From the above Hamiltonian four-momentum of the charged particle has the form   
\begin{eqnarray}
p^{\mu}=g^{\mu\nu}\left(\pi_{\nu}-qA_{\nu}\right)\, . 
\end{eqnarray}
Then the equations for the charged particle motion can be written as 
\begin{eqnarray} 
\label{Eq:H1}
  \frac{dx^\alpha}{d\lambda} &=& \frac{\partial H}{\partial \pi_\alpha}\, ,  \\
  \frac{d\pi_\alpha}{d\lambda} &=& - \frac{\partial H}{\partial x^\alpha}\, , \label{Eq:H2}
\end{eqnarray}
with $\lambda=\tau/m$ being the affine parameter associated with the proper time $\tau$ for
timelike geodesics. }

\textcolor{black}{From Eqs.~(\ref{Eq:H1}) and (\ref{Eq:H2}) it is then  straightforward to obtain the equations of motion for the charged particle. Note that
the magnetized black hole spacetime considered admits two Killing vectors, $\xi^{\mu}_{(t)}=(\partial/\partial t)^{\mu}$ and
$\xi^{\mu}_{(\phi)}=(\partial/\partial \phi)^{\mu}$ being responsible for a stationary and an axisymmetry, 
and thus there appear two conserved quantities, the energy and angular momentum of the charged particle together with $m^2$ (mass of the charged particle). However, we will not consider the other one related to the latitudinal motion as we restrict ourselves to the equatorial motion, i.e. $\theta=\pi/2$. Following Eqs.~(\ref{Eq:H1}) and (\ref{Eq:H2}), the first two can be determined by the following equations
\begin{eqnarray}
\label{En1} \pi_t-qA_{t}&=& -g_{\mu\nu}(\xi_{t})^{\mu}p^{\nu}=
g_{tt}p^{t} + g_{t\phi}p^{\phi}\ ,\\
\nonumber\\
 \label{Ln1}
\pi_{\phi}-qA_{\phi}&=& -g_{\mu\nu}(\xi_{\phi})^{\mu}p^{\nu}=g_{\phi t}p^{t} +
g_{\phi\phi}p^{\phi}\, ,
\end{eqnarray}
with the quantities $\pi_t\equiv -E$ and $ \pi_{\varphi} \equiv L $ being the constants of motion.}

%%%%%%%%%%%%%%%%%%%%%%%%%%%%%%%%%%%%%%%%%%%%%%%%%%%%%%%%%5
\textcolor{black}{Using Eqs. (\ref{Eq:H1}), (\ref{Ln1}), and the normalization condition $g_{\mu\nu}p^{\mu}p^{\nu}=-m^2$}, one can then easily write the radial part of the equation of motion for the massive test particle in the equatorial plane (i.e. $\theta=\pi/2$) in the following form 
\begin{eqnarray}
\frac{1}{2}\dot{r}^{2} + R(r,\mathcal{L},\mathcal{E}) &=&0,
\end{eqnarray}
where the function $ R(r,\mathcal{L},\mathcal{E})$ governing the radial motion is given by
\begin{widetext}
\begin{eqnarray}\label{Eq:Effective}
R(r,\mathcal{L},\mathcal{E})&=&\frac{fH^-1 }{ \Big(f H^2+r^2 (\omega -1) \omega \Big)r^2 }\bigg[-f H^2 \Big(H \mathcal{L}^2+ r^2\Big)+ H q \bigg(A_{\phi } \Big[q A_{\phi } \left(r^2 \omega -f H^2\right)+2 f H^2 \mathcal{L}+ 2 r^2 \omega  (\mathcal{E} -\mathcal{L})\Big] \nonumber\\ &+&  2 r^2 A_t \left(q \omega  A_{\phi }+\mathcal{E}-\omega  \mathcal{L}\right)+q r^2 A_t^2\bigg)+r^2 \bigg(H \left(\mathcal{E}^2+\omega  \mathcal{L}^2-2 \mathcal{E}  \omega  \mathcal{L}\right)- r^2 (\omega -1) \omega \bigg)\bigg]\, ,
\end{eqnarray}
\end{widetext}
where we have used specific conserved quantities per unit mass, i.e., $\mathcal{E}=E/m$ and
$\mathcal{L}=L/m$ representing the energy and angular momentum of the massive particle. We are studying timelike nongeodetic motion for which the radial motion is determined by the following equation,
\begin{eqnarray}
\dot{r}^{2} = \left[\mathcal{E} -
\mathcal{E_{+}}(r)\right]\left[\mathcal{E}
-\mathcal{E_{-}}(r)\right] = 0 \, .
\end{eqnarray}
It is clear that $\dot{r}^2\geq 0$ and hence we have either $\mathcal{E}>\mathcal{E}_{+}(r)$ or $\mathcal{E}<\mathcal{E}_{-}(r)$. However, we will restrict ourselves to the positive energy $\mathcal{E}_{+}(r)=V_{eff}(r)$ for future pointing physically meaningful timelike motion. \textcolor{black}{Note that in the above equation $\mathcal{E}_{-}(r)$ and $\mathcal{E}_{+}(r)$ are negative and positive solutions of the effective potential.} The effective potential for radial motion of the charged particle moving around a magnetized black hole is then given by
\begin{eqnarray}
V_{eff}(r)&=&\frac{\Big[H q A_{\phi } \left(q A_{\phi }-2 \mathcal{L}\right)+H \mathcal{L}^2+r^2\Big]^{1/2}}{H^{1/2}\, r\, \Big[f H^2+r^2 (\omega -1) \omega \Big]^{-1/2} }\nonumber\\&-& q A_t+\omega  \left(\mathcal{L}-q A_{\phi }\right)\, ,
\end{eqnarray}
where the metric functions $f$, $H$, and $\omega$ are given in Eq.~(\ref{Eq:fhw}) while the electromagnetic four-vector potentials $A_{t}$ and $A_{\phi}$ in Eq.~(\ref{Eq:4-vec}). 
Note that the effective potential in the above reduces to the Reissner-Nordstrom case for $B\to 0$, while it reduces to the Schwarzschild spacetime for $B,Q\to 0$.
\begin{figure*}
\centering
\includegraphics[width=0.45\textwidth]{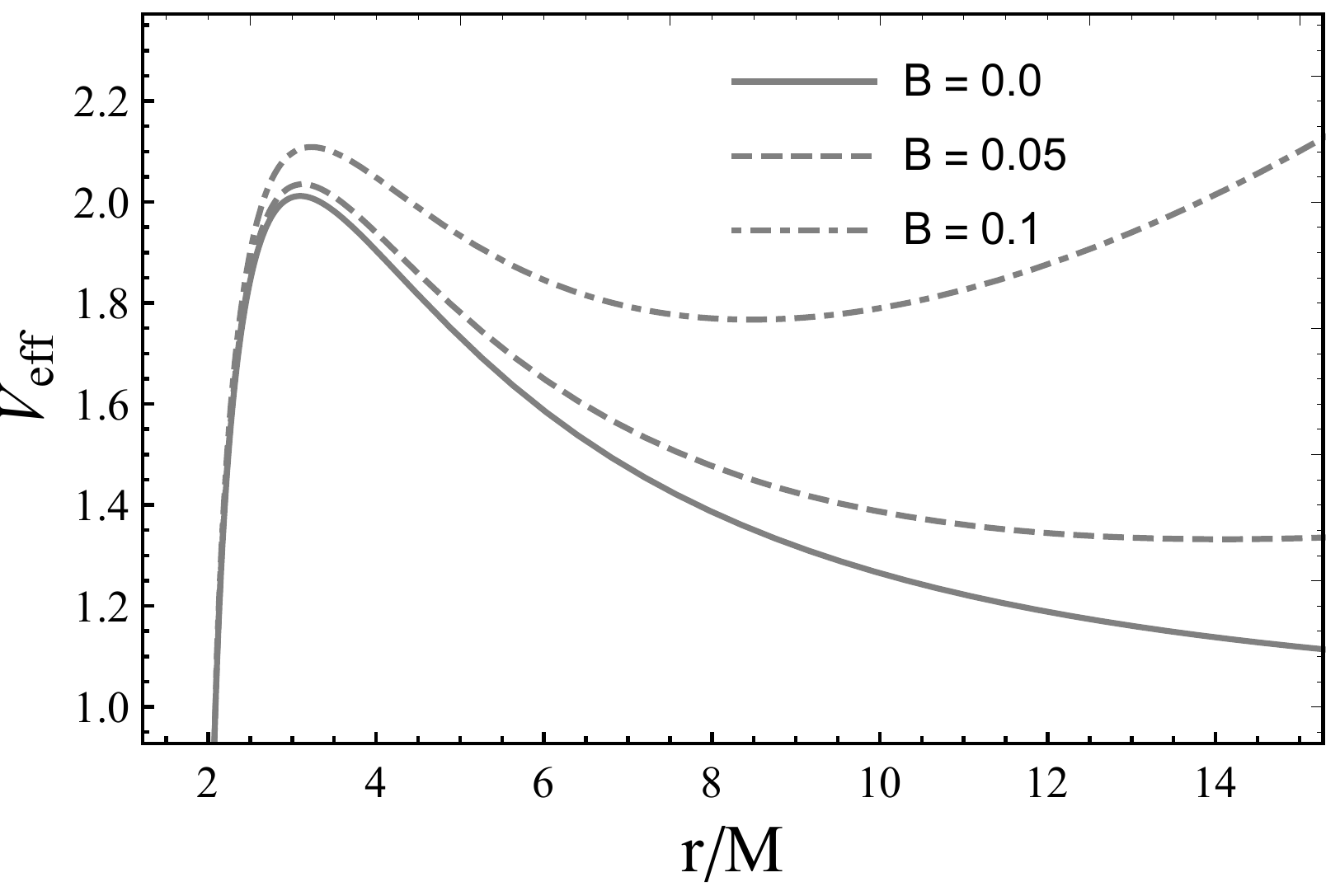}
\includegraphics[width=0.45\textwidth]{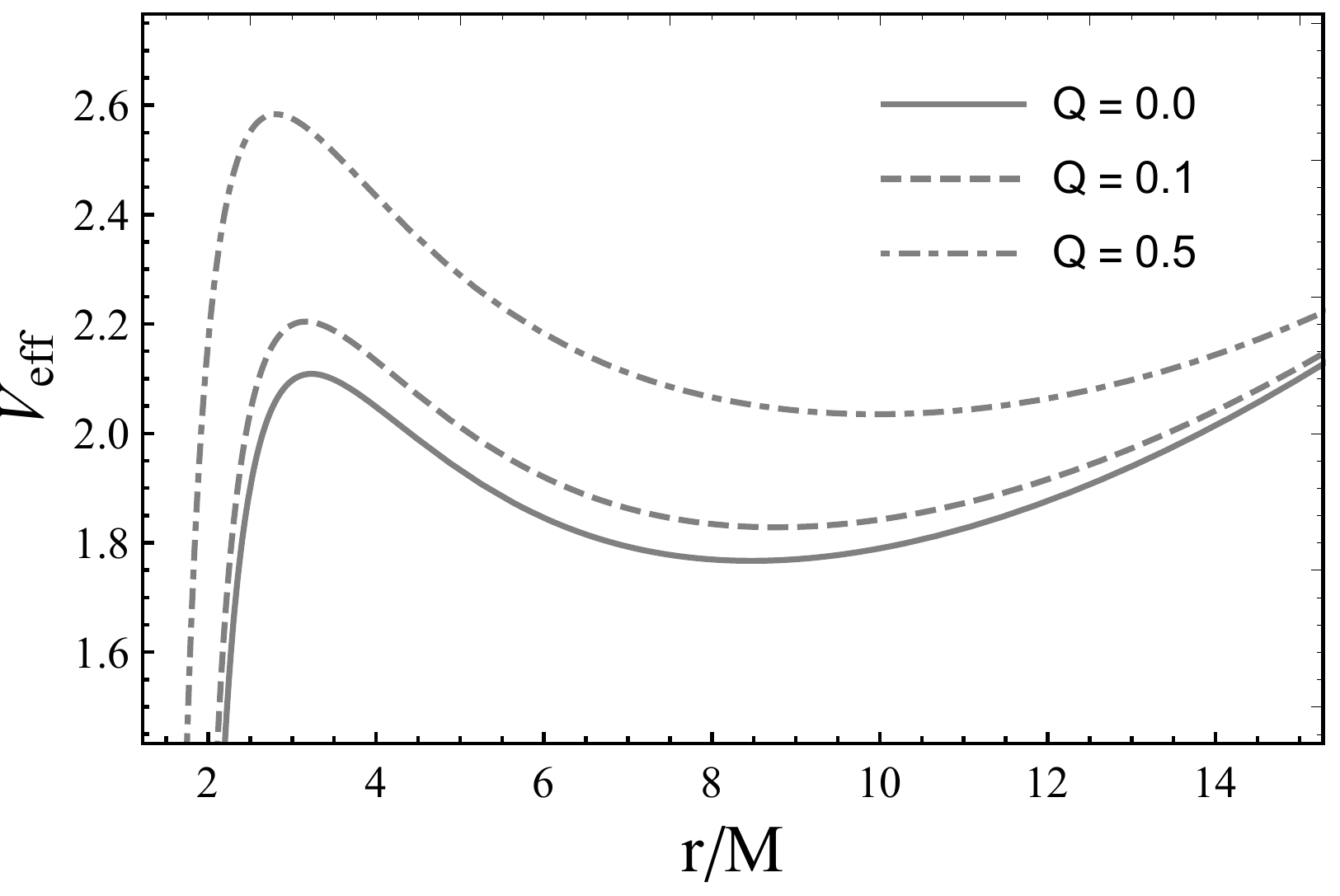}

\caption{\label{fig:eff1} Radial dependence of the effective potential for the radial motion of massive neutral particle around the magnetized black hole. Left panel shows $V_{eff}$ for different values of magnetic field $B$ in the case of $Q=0$, while the right panel shows $V_{eff}$ for different values of black hole charge $Q$ in the case of fixed $B=0.1$. }
\end{figure*}
\begin{figure*}
\centering
\includegraphics[width=0.45\textwidth]{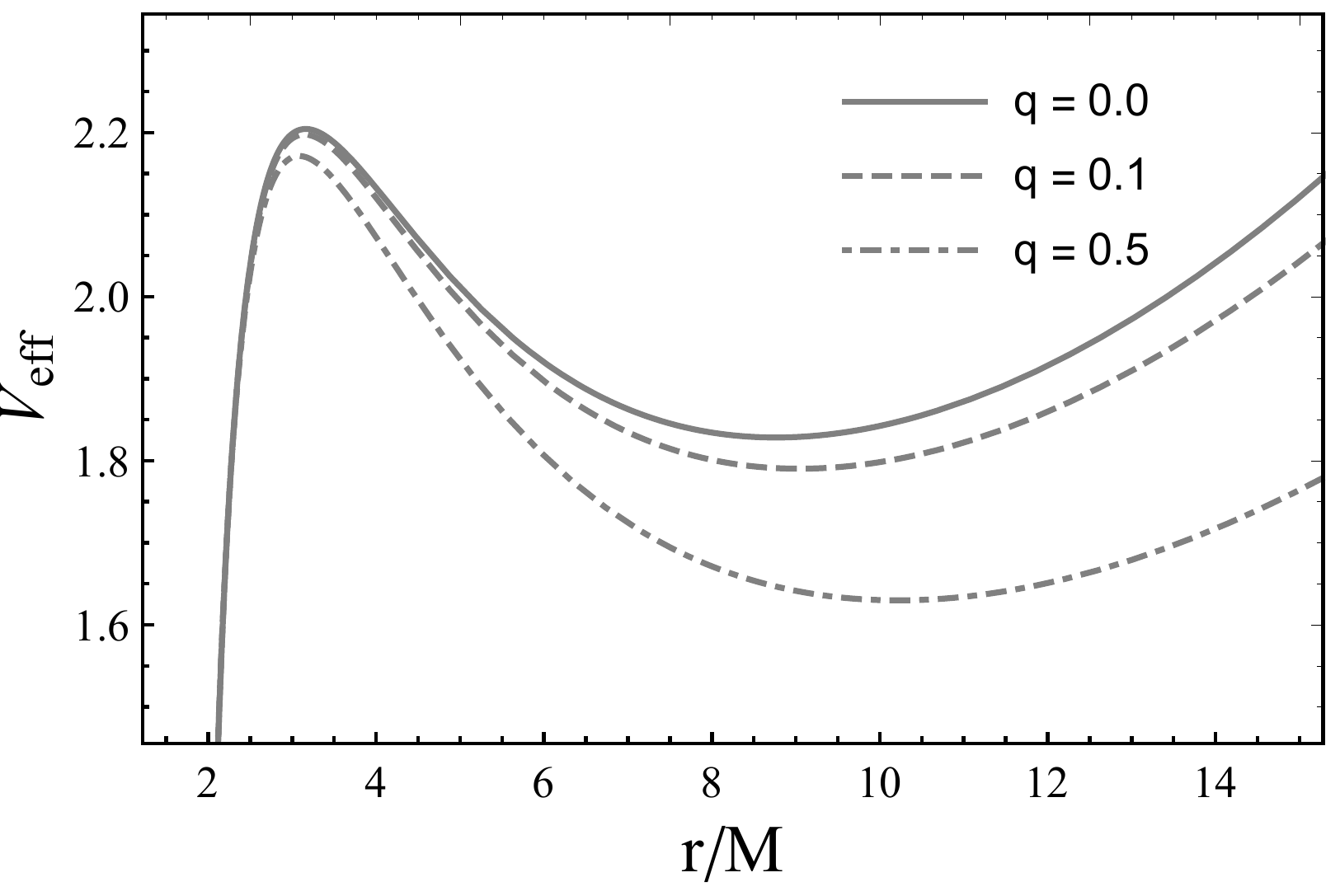}
\includegraphics[width=0.45\textwidth]{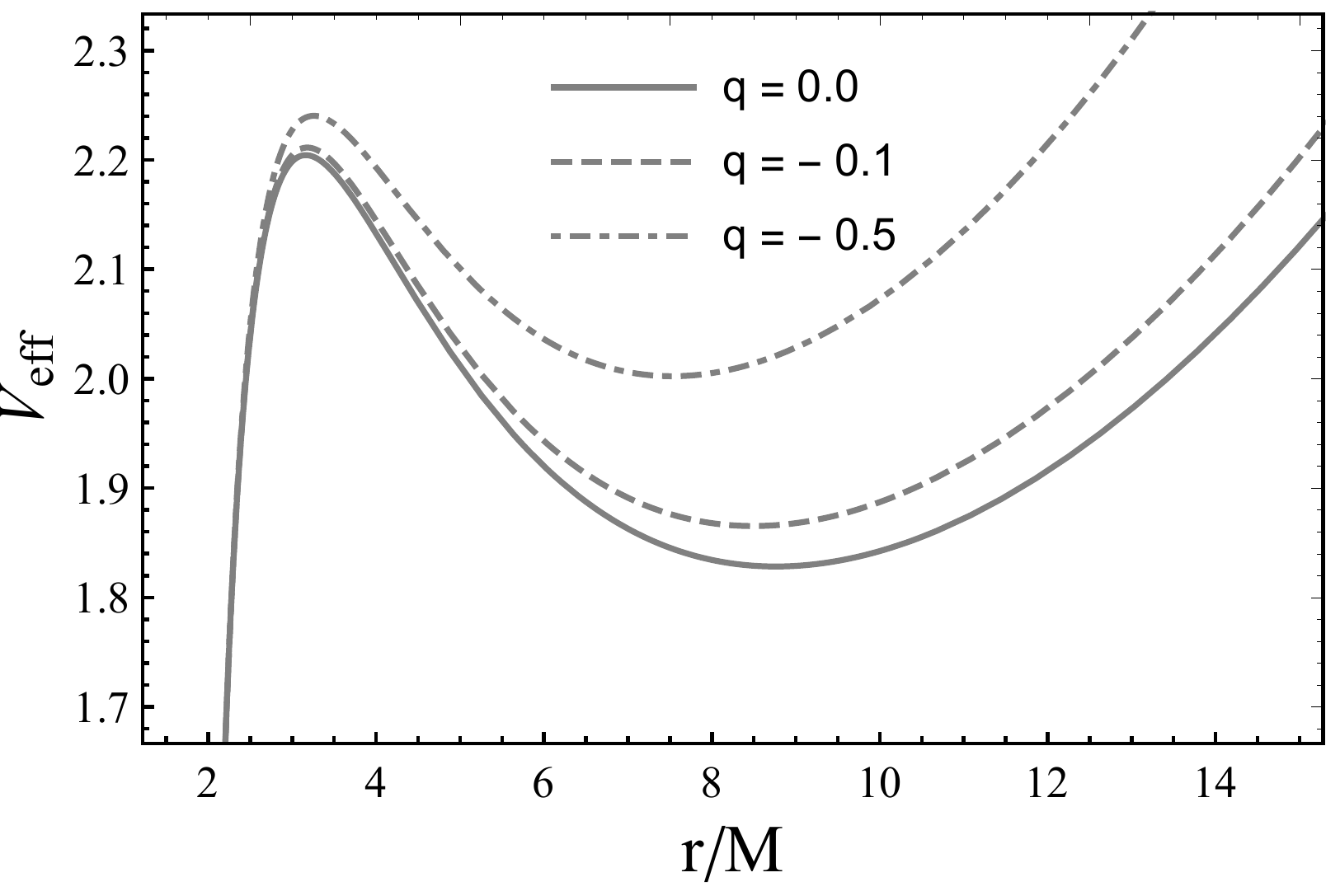}

\caption{\label{fig:eff2} Radial dependence of the effective potential for the radial motion of massive charged particle around the magnetized black hole. Left panel shows $V_{eff}$ for different values of positive charge $q>0$, while the right panel shows $V_{eff}$ for different values of negative charge $q<0 $ in the case of fixed $Q=0.1$ and $B=0.1$. }
\end{figure*}
\begin{figure*}
\centering
 \includegraphics[width=0.3\textwidth]{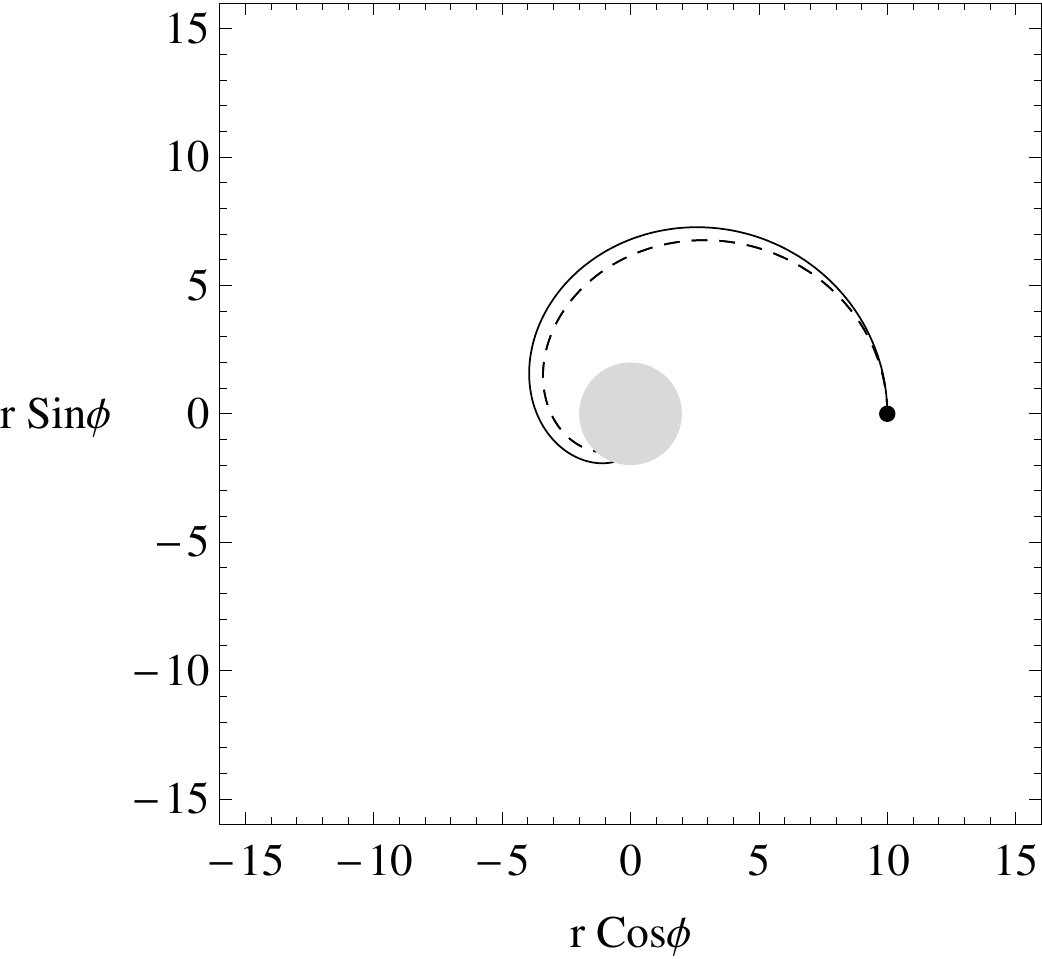}% Here is how to import EPS art
 \includegraphics[width=0.3\textwidth]{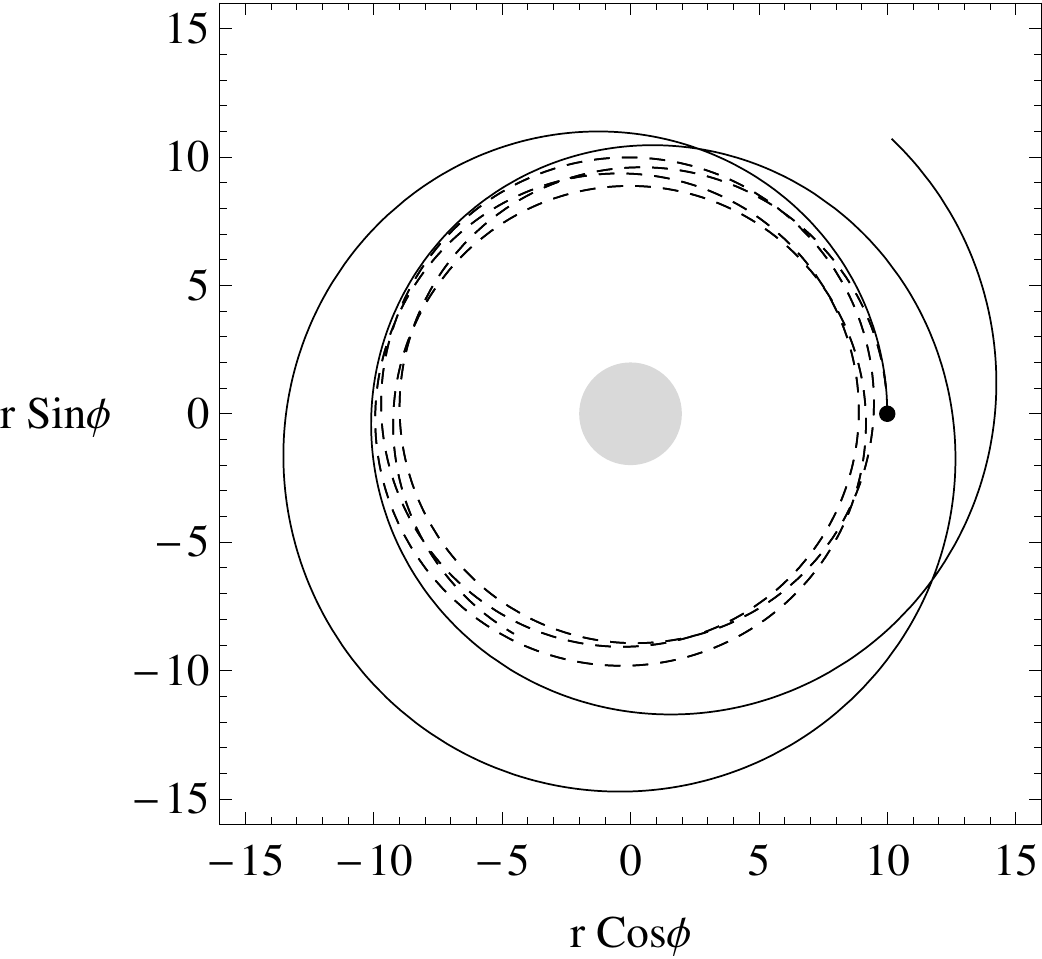}
 \includegraphics[width=0.3\textwidth]{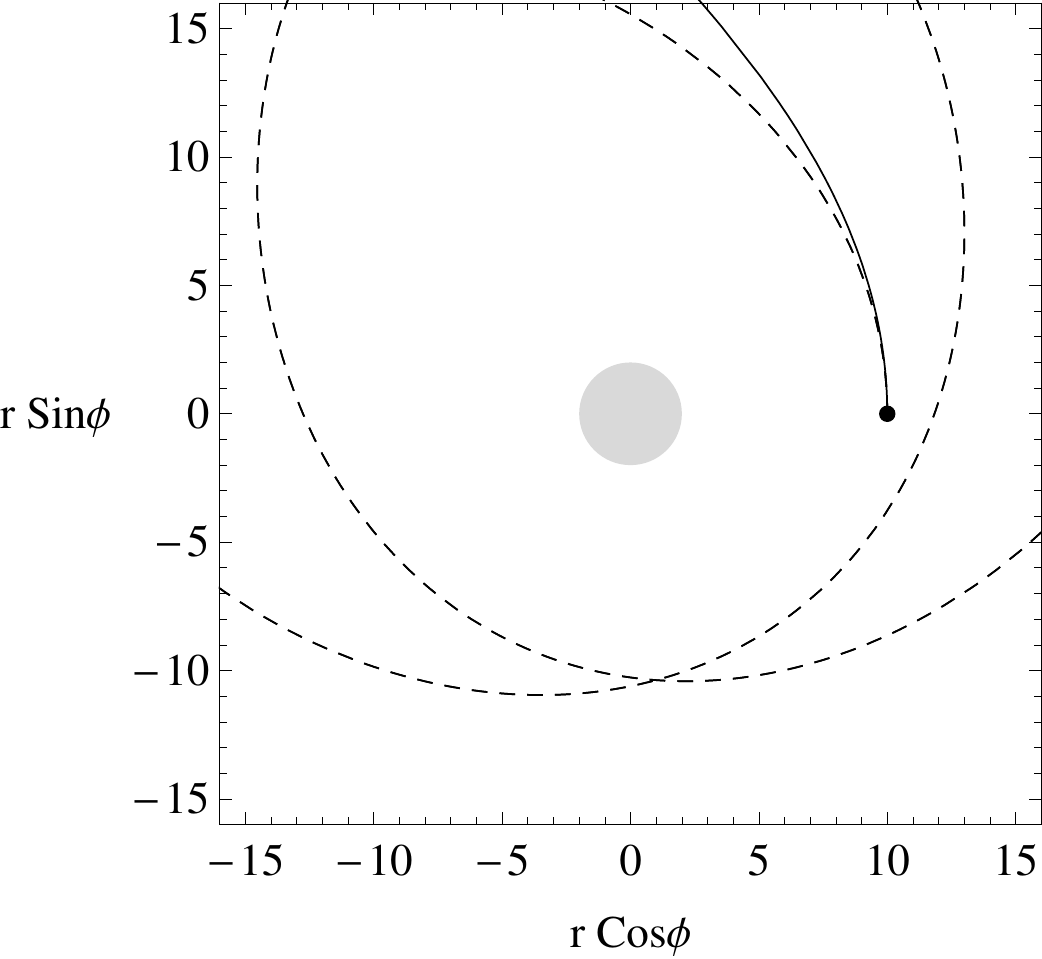}% Here is how to import EPS art
\caption{\label{fig:traj1} Trajectories of the massive particle in the equatorial plane of the magnetized black hole for $B=0$ (solid curve) and $B=0.01$ (dashed curve).  Left/middle/right panels respectively refer to $\mathcal{L}=3,4,6$ for all possible orbits. %Note that particle starts from $r_{0}=10$ towards the black hole.
Here, in all plots black hole electric charge is taken to be $Q=0$.}
\end{figure*}
\begin{figure*}
\centering
\includegraphics[width=1.0\textwidth]{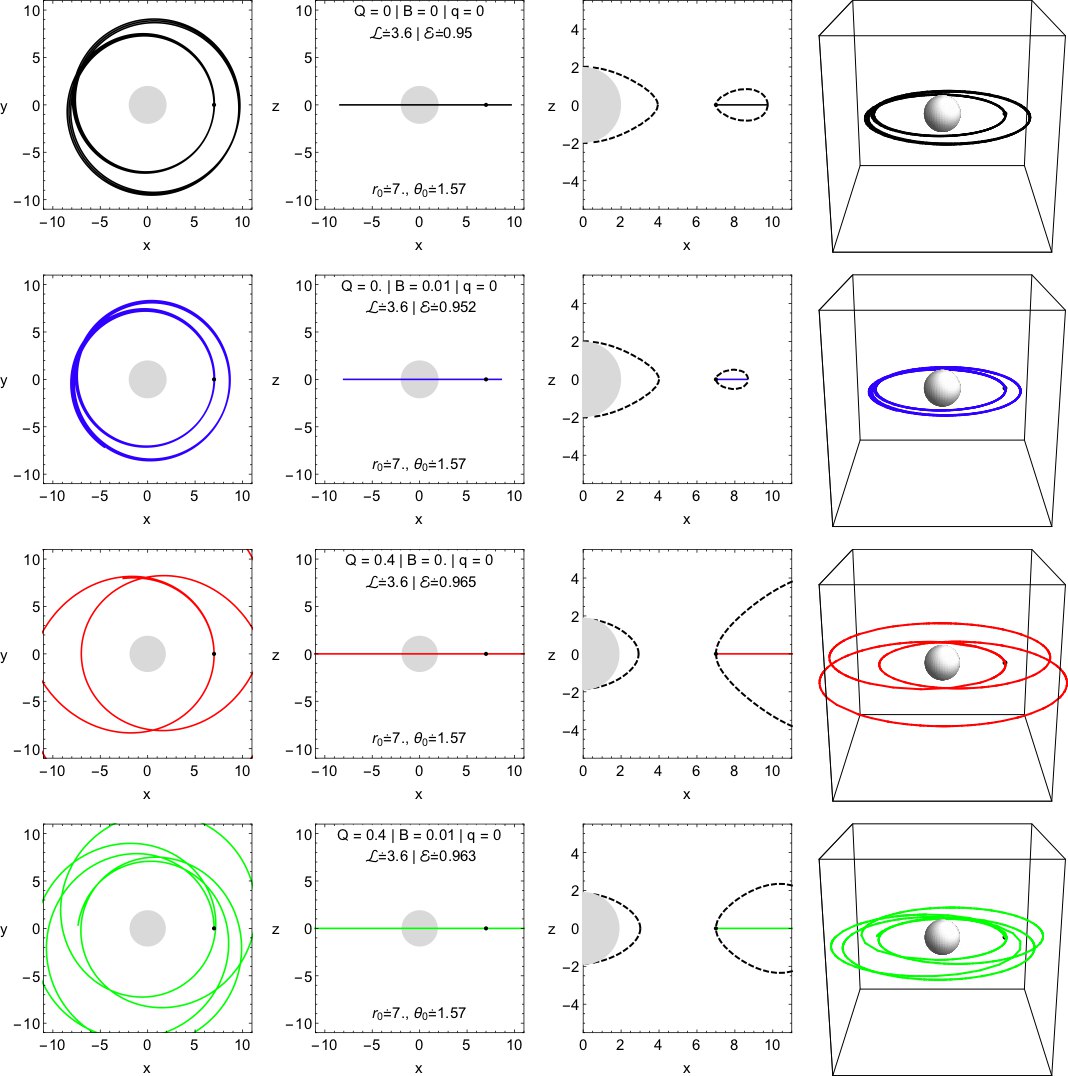}%

\caption{\label{fig:traj2} Trajectories of neutral particle in the equatorial plane of the magnetized black hole for given values of black hole charge $Q$ and magnetic field parameter $B$ for various cases. %Note that particle starts from $r_{0}=7$ towards the black hole.  
}
\end{figure*}
\begin{figure*}
\centering
\includegraphics[width=1.0\textwidth]{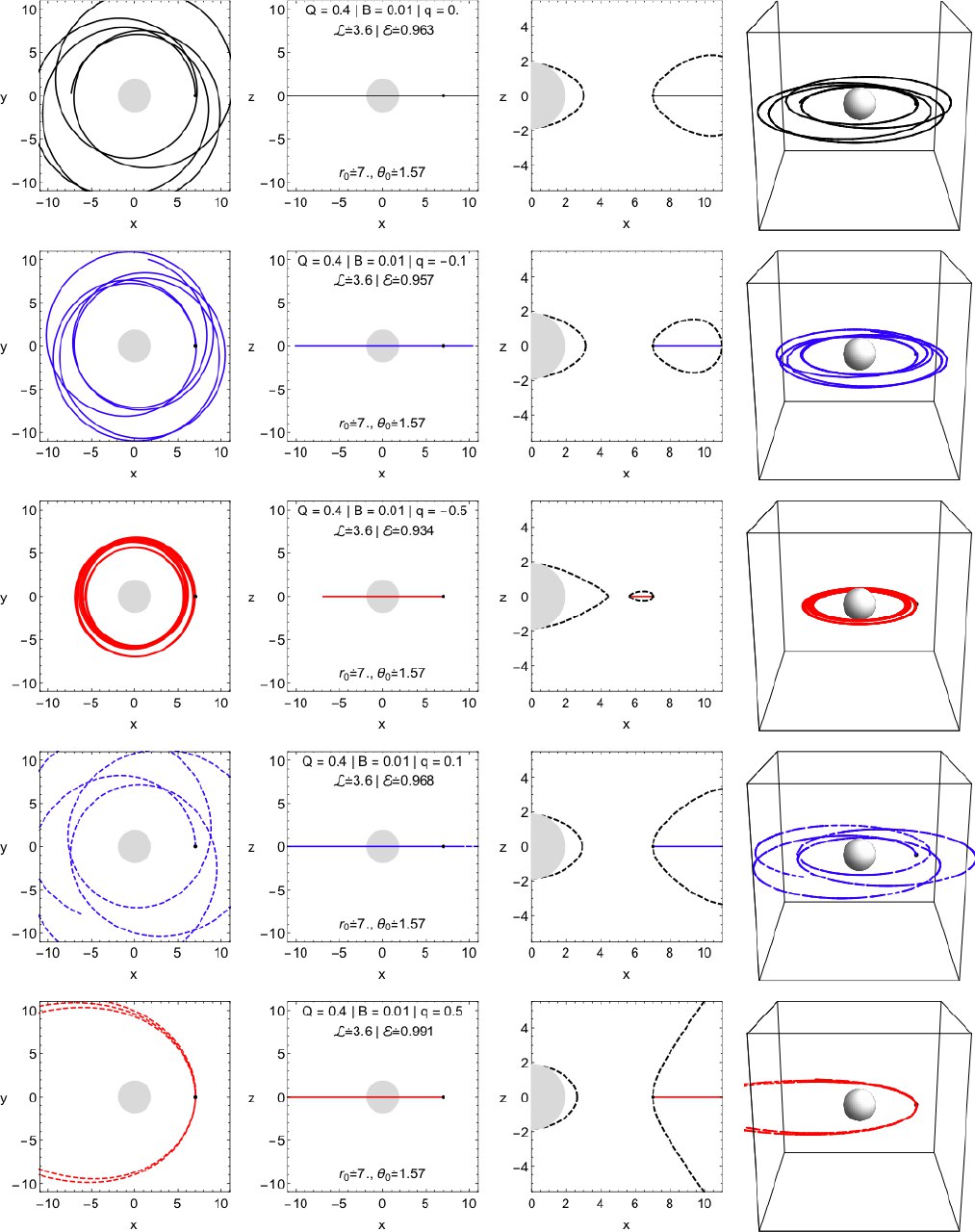}% 

\caption{\label{fig:traj3} Trajectories of the charged particle in the equatorial plane of the magnetized black hole for different values of both negative and positive particle charge $q$ in the case of fixed $Q$ and $B$. %Note that the charged particle starts from $r_{0}=7$ towards the black hole.
} 
\end{figure*}

 Now we analyze $V_{eff}$ for nongeodesic motion of test particles in the gravitational and electromagnetic fields of magnetized black hole. In Fig.~\ref{fig:eff1}, we show the effective potential for the radial motion of neutral particle moving around magnetized black hole. In Fig.~\ref{fig:eff1}, the left panel shows the impact of magnetic field $B$ on the radial profile of the effective potential for $Q=0$, while the right panel shows the impact of the black hole electric charge for fixed $B=0.1$. As shown in Fig.~\ref{fig:eff1}, the strength of the effective potential increases as the magnetic field $B$ increases, while the shape of the effective potential shifts left to smaller $r$ with increasing $Q$. The orbits of neutral particles under the combined effects of magnetic field $B$ and black hole electric charge $Q$ can become close to the black hole horizon as seen in Fig.~\ref{fig:eff1}. In Fig.~\ref{fig:eff2},  the left panel reflects the impact of the positive charge $q>0$ on the radial profile of the effective potential, while the right panel reflects the impact of the negative charge $q<0$ in the case of fixed $B=0.1$ and $Q=0.1$. As can be seen from Fig.~\ref{fig:eff2}, the orbit of the charged particle becomes close to smaller $r$ as negative charge $q<0$ increases while it shifts toward right to larger $r$ with increasing positive charge $q>0$. 

Let us then turn to consider particle trajectory in the vicinity of magnetized black hole. For further analysis of particle trajectory, we shall for simplicity restrict motion to the equatorial plane, i.e. $\theta=\pi/2$. Fig.~\ref{fig:traj1} reflects the role of magnetic field $B$ on the trajectory of neutral particle in the equatorial plane for different values of angular momentum $\mathcal{L}$. As shown in Fig.~\ref{fig:traj1}, this clearly shows that the trajectory of particles becomes close to the central object as a consequence of the presence of the magnetic field parameter $B$, as also seen in Fig.~\ref{fig:eff1}. Thus, the magnetic field parameter $B$ of magnetized black hole acts as a gravitational repulsive charge like a black hole electric charge. Further we also give detailed plots representing particle trajectory as a consequence of the presence of black hole electric charge $Q$ and magnetic field parameter $B$ for both neutral and charged particles, see Figs.~\ref{fig:traj2} and \ref{fig:traj3}.   
\begin{figure*}

  \includegraphics[width=0.3\textwidth]{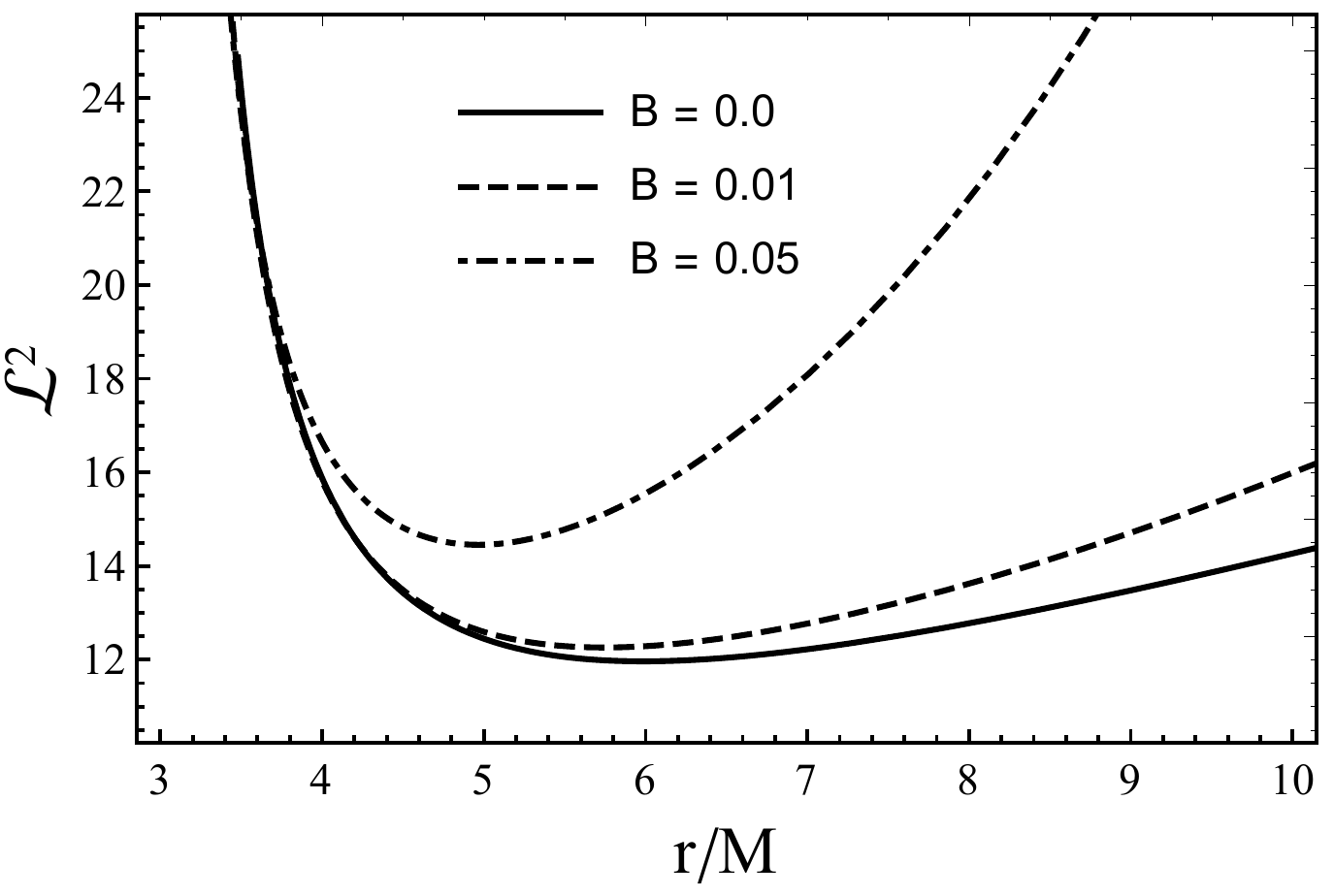}% 
  \includegraphics[width=0.3\textwidth]{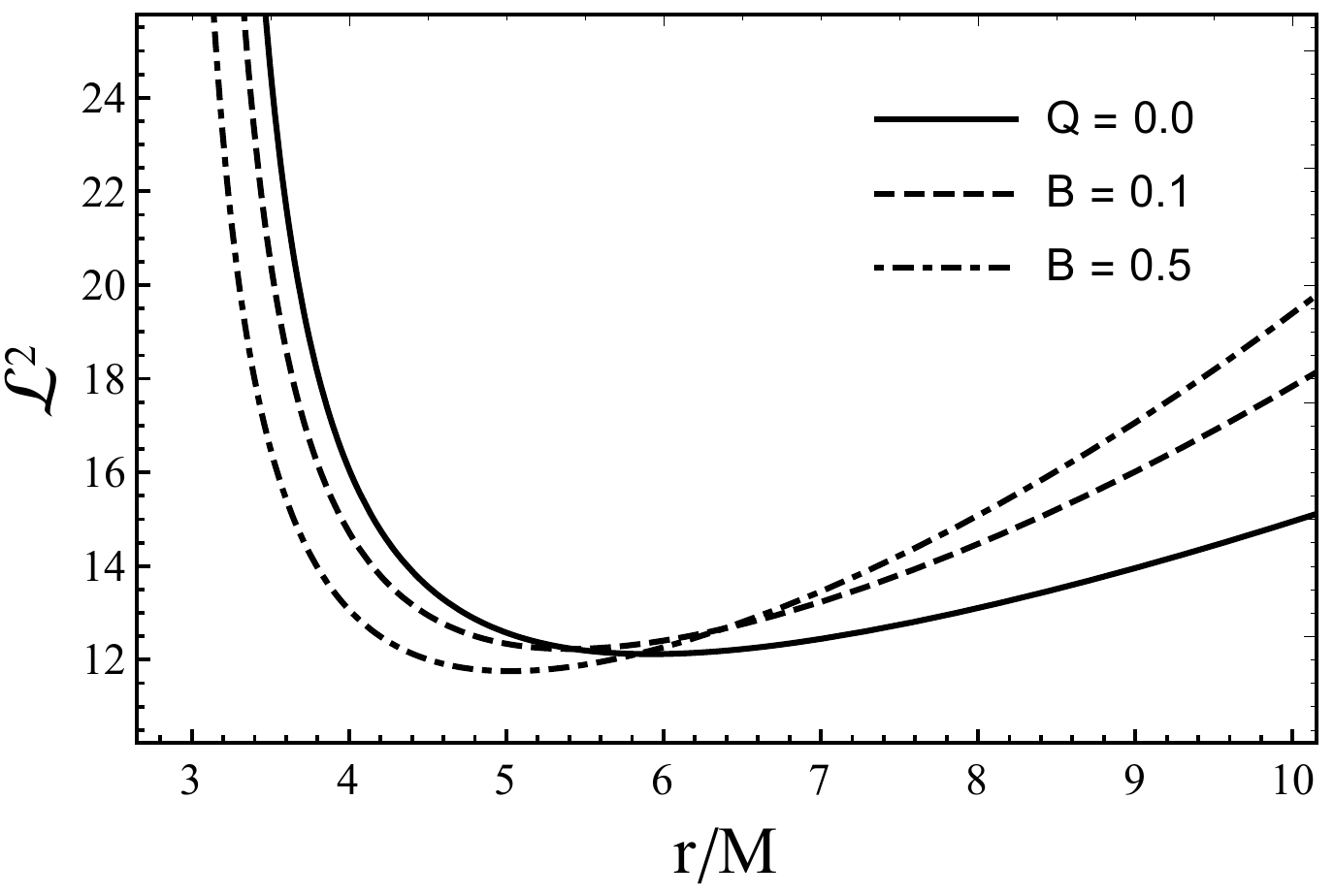}
  \includegraphics[width=0.3\textwidth]{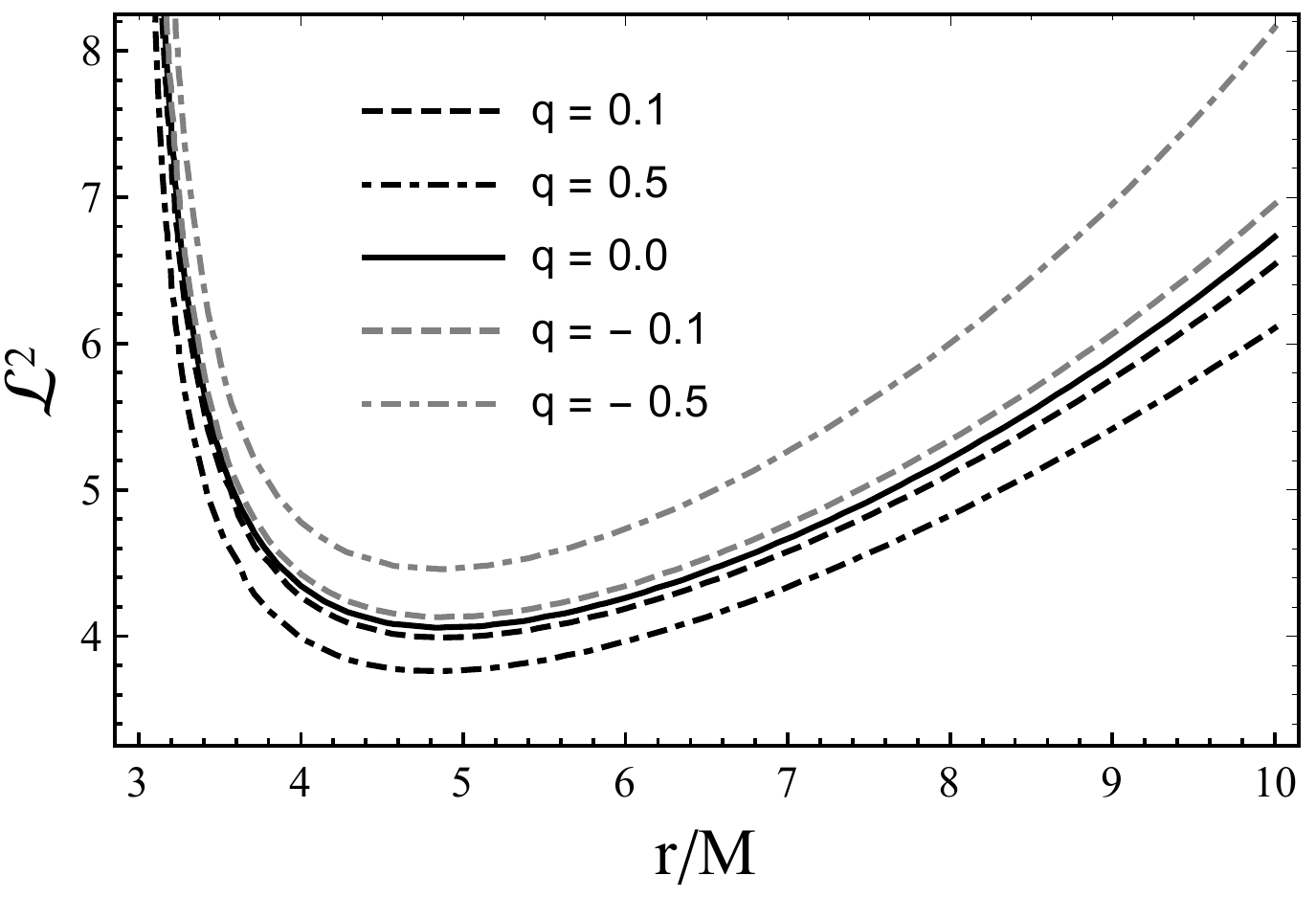}
\caption{\label{fig:ang} The radial dependence of the angular momentum for test particles orbiting around a magnetized Reissner-Nordstr\"{o}m black hole with magnetic charge $q=0$. Left panel:  ${\mathcal L}^2$ is plotted for different values of $B$ for fixed $Q=0.1$. Middle panel: ${\mathcal L}^2$ is plotted for different values of $Q$ for fixed $B=0.01$. Right panel: ${\mathcal L}^2$ is plotted for different values of both negative and positive values of particle charge $\pm q$ in the case of fixed $Q=0.1$ and $B=0.05$. }
\end{figure*}
\begin{table}[h]
\caption{\label{1tab} The values of the ISCO radius ($r_{i}$), the angular momentum ($\mathcal{L}_{i}$) and the energy ($\mathcal{E}_{i}$) of the neutral test particles orbiting on the ISCO around the magnetized Reissner-Nordstr\"{o}m black hole are tabulated for different values of $Q$ and $B$. }
\begin{ruledtabular}
\begin{tabular}{c|cccccc}
 &  &  &  &  $B$ &  &   \\ \hline
{$\rm Q $} & & $0.00$ & $0.01$ &  $0.04$
& $0.08$ & $0.10$ \\
\hline
 0.0 & $r_{i}$ & 6.00000 &5.91997 & 5.32285 & 4.75062 & 4.56610  \\
   & $\mathcal{L}_{i}$  & 3.46410 &3.48144 & 3.70022 & 4.28076 & 4.72620 \\
    & $\mathcal{E}_{i}$   & 0.94281 &0.94505 & 0.97676 & 1.07860 & 1.16614 \\\\
0.1  & $r_{i}$ &5.98497  &5.74518  &5.13935  &4.68373 &4.52912  \\
 & $\mathcal{L}_{i}$ &3.45928  &3.52444  &3.81248  &4.35519 &4.72863  \\
& $\mathcal{E}_{i}$ &0.94267  &0.95548  &1.01262  &1.13196 &1.21873  \\\\
0.2  &$r_{i}$  &5.93957  &5.57969 &4.97438 &4.60136 &4.47579  \\
 & $\mathcal{L}_{i}$  &3.44471  &3.55248 &3.89573 &4.40613 &4.72071  \\
& $\mathcal{E}_{i}$  &0.94227  &0.96522 &1.04589 &1.18113 &1.26735  \\\\
0.3  &$r_{i}$ &5.86278   &5.41358 &4.81505 &4.50213 &4.40397  \\
&$\mathcal{L}_{i}$  &3.42006   &3.56508 &3.95200 &4.43514 &4.70207  \\
&$\mathcal{E}_{i}$ &0.94158   &0.97416 &1.07653 &1.22626 &1.31222  \\\\
0.5 &$r_{i}$ &5.60664   &5.05352 &4.48236 &4.24591 &4.19363 \\
&$\mathcal{L}_{i}$ &3.33774   &3.53997 &3.98542 &4.42783 &4.62776 \\
&$\mathcal{E}_{i}$ &0.93916   &0.98911 &1.12921 &1.30415 &1.39062 \\\\
1.0 &$r_{i}$ &4.00000   &3.57336 &3.16038 &2.99142 & 2.96795 \\
&$\mathcal{L}_{i}$ &2.82843   &3.03584 &3.45352 &3.83301 & 3.98326 \\
&$\mathcal{E}_{i}$ &0.91856   &0.99168 &1.17878 &1.38671 & 1.47824 \\
\end{tabular}
\end{ruledtabular}
\end{table}
\begin{table}[h]
\caption{\label{2tab} The values of the ISCO radius $r_{i}$ are tabulated in the case of the charged particle orbiting around the magnetized Reissner-Nordstr\"{o}m black hole for various values of charge $\pm q$ for the fixed $B=0.1$.  }
\begin{ruledtabular}
\begin{tabular}{c|cccc}
 &  &  $q$ &  &   \\ \hline
{$\rm Q $}  & $0.01$ &  $0.05$
& $0.10$ & $0.50$  \\
     & $-0.01$ &  $-0.05$ & $-0.10$ & $-0.50$  \\
\hline
0.1    &4.52982  &4.53219  &4.53413 &4.49620   \\
       &4.52838  &4.52504  &4.52009  &4.46278  \\\\
0.2    &4.47615 &4.47708  &4.47706 &4.41627  \\
       &4.47539 &4.47335 &4.46989  &4.42156  \\\\
0.3    &4.40338 &4.40295 &4.40045 &4.31340  \\
       &4.40401 &4.40367 &4.40222 &4.36613  \\ \\
0.5    &4.19223 &4.18596 &4.17649 &4.02520  \\
       &4.19496 &4.19964 &4.20417 &4.20550  \\ \\
1.0    &2.95865 &2.92014 &2.86902 &2.30516  \\
       &2.97713 &3.01258 &3.05416 &3.29255  \\
%
% {$ B $}      & $0.01$ &  $0.04$ & $0.08$ & $0.10$  \\ 
 %\hline        &   &  $B$ &  &   \\
%      
\end{tabular}
\end{ruledtabular}
\end{table}

Now we focus on the study of the circular orbits of test particles around the magnetized Reissner-Nordstr\"{o}m black hole.  For test particles to be on the circular orbits one needs to solve the following standard equations
\begin{eqnarray}\label{Eq:circular}
V_{eff}(r)=\mathcal{E}\,  \mbox{~~and ~~}  \frac{\partial V_{eff}(r)}{\partial r}=0\, .
\label{Eq:cir2}
\end{eqnarray}
By solving above equations we can then evaluate the values of the angular momentum $\mathcal{L}$ and energy $\mathcal{E}$ for which stable circular orbits are allowed for test particles. In Fig.~\ref{fig:ang}
we show the radial profiles of the angular momentum for test particles on the circular orbit. One can see from Fig.~\ref{fig:ang} that circular orbits shift toward left to small radii $r$ with an in increase in the value of both $Q$ and $B$, thus reducing the radii of circular orbits. However, both $Q$ and $B$ have similar effect, the value of $\mathcal{L}$ increases with increasing $B$, while it decreases with increasing $Q$ for test particle to be on circular orbit, as seen in Fig.~\ref{fig:ang}. For the charged particle the value of $\mathcal L$ depends on the sign of test particle electric (positive and negative) charge (see Fig.~\ref{fig:ang}, right panel). Finally, to determine the radius of innermost stable circular orbit (ISCO) of test particles around the magnetized Reissner-Nordstr\"{o}m black hole we need to solve the following equation 
\begin{eqnarray}\label{Eq:isco}
\frac{\partial^2 V_{eff}(r)}{\partial r^2}=0\, .
\end{eqnarray}
In Tables~\ref{1tab} and \ref{2tab}, we explicitly show the ISCO radius of neutral and (positively and negatively) charged particles as a consequence of solving Eqs.~(\ref{Eq:circular}) and (\ref{Eq:isco}) numerically for different values of black hole electric charge $Q$ and magnetic field $B$. In Table~\ref{1tab} we also show the value of angular momentum $\mathcal{L}$ and energy $\mathcal{E}$ at the ISCO.  As shown in Table~\ref{1tab} this clearly shows that for neutral particle the ISCO radius decreases as a consequence of presence of black hole electric charge $Q$ and magnetic field $B$. For the charged particle the ISCO radius increases (decreases) as a consequence of presence of positive (negative) particle electric charge, as seen in Table~\ref{2tab}.  We notice that the value of ISCO radius however gets radically altered for large values of black hole electric charge $Q$. This happens because the repulsive (attractive) Coulomb force for positively (negatively) charged particle turns out to be greater than the Lorentz force arising from the magnetic field, see Table~\ref{2tab}.

\begin{figure}

 \includegraphics[width=0.45\textwidth]{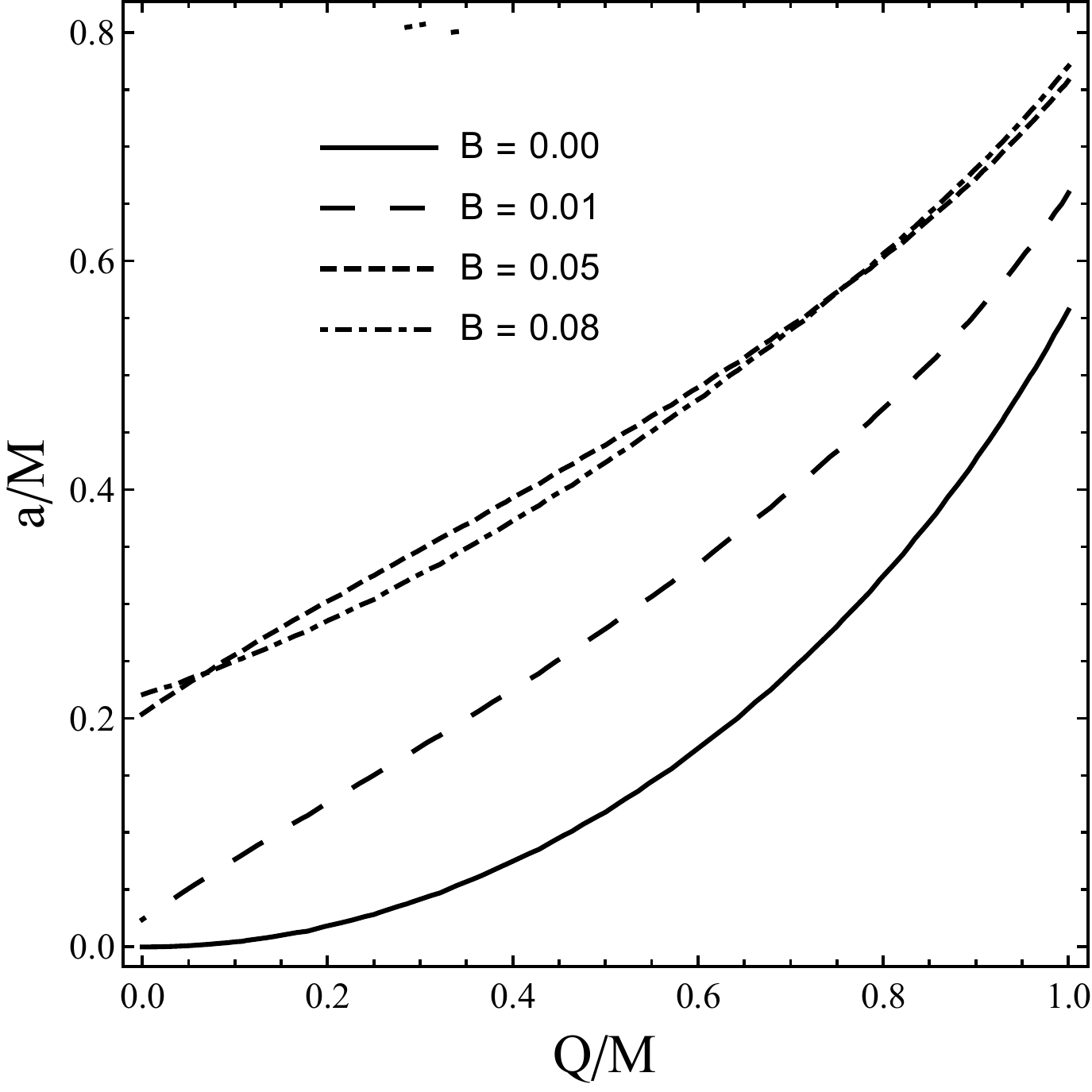}
 
\caption{\label{fig:mimic} The plot shows the values of spin parameter $a$ as a function of black hole charge $Q$ for which the degeneracy appears for value of the ISCO. For a given value of $B$ the Kerr and magnetized Reissner-Nordstr\"{o}m black hole geometries have the same ISCO.} \end{figure}

Let us then turn to the study of comparison between Kerr and magnetized Reissner-Nordstr\"{o}m black hole geometries. It is well known that the rotating black hole causes axially symmetric spacetime. Similarly, the presence of magnetic field turns a spherical symmetric spacetime into axially symmetric one. However, distance observer is not able to distinguish between two geometries using astronomical observations. Although much progress has been made in observational studies, the LIGO and Virgo scientific collaborations \cite{Abbott16a,Abbott16b} and the Event Horizon Telescope (EHT) collaboration~\cite{Akiyama19L1,Akiyama19L6} have not shown explicit departures about black hole parameters yet except its total mass $M$. Thus, black holes have been still referred to as the best candidates. Thus, one can make predictions for black hole parameters in accordance with these facts. With this motivation, we then focus on the value of black hole electric charge $Q$ and magnetic field $B$. In Fig.~\ref{fig:mimic} we show the values of spin parameter $a$ as a function of black hole charge $Q$ for different values of black hole magnetic field. One can easily see that for a given value of $B$ the Kerr and magnetized Reissner-Nordstr\"{o}m black hole geometries have the same ISCO.  As shown in Fig.~\ref{fig:mimic}, black hole charge and magnetic field can mimic black hole spin $a$ up to $a/M\approx 0.8$. As a consequence of this result, analyzing electromagnet radiations emitted by accretion disk could give the same observations about two black hole geometries being observed.  This would of course be a theoretical model, it does however play an important role as our main purpose is to study a whole process and the qualitative aspects of the magnetized Reissner-Nordstr\"{o}m black hole geometry. 

\section{Particle collisions}\label{Sec:4}

Black holes are known as one of the most energetic astrophysical gravitational objects in the Universe. Energy extraction from a black hole can be explained with the use of different mechanisms such as Penrose processes and Blandford-Znajek mechanism (see for example \cite{wagh85, Hawking74, Hawking75, Hawking76, wagh85a}). Authors of the work \cite{Banados09} have demonstrated that the center of mass energy (CM) of colliding particles goes up exponentially in the vicinity of Kerr black hole extracting some fraction of the rotational energy of a black hole. Therefore, it is worth studying particle collision near black as it allows to get information about the energetic properties of such compact objects. In this section, we focus on some simple scenarios of particle collisions around the magnetized Reissner-Nordstr\"{o}m black hole that may serve as a particle accelerator to high energy phenomena. 

\subsection{Collisions of neutral particles with opposite angular momentum \label{subsect1}}

Here we evaluate the CM energy of colliding test particles having zero electric charge and falling with zero velocity from infinity to the central black hole, as described by the line element given in Eq.~(\ref{Eq:metric}). We assume that these two particles have opposite angular momenta $L_1=-L_2=L$ and the same initial energies $E_{1,2}=m$ at infinity. For that, the four-velocity of test particles can be written as

\begin{eqnarray}
\nonumber \label{conU}
u_i^t&=&-g^{tt} + g^{t\phi} \frac{L_i}{m} \ ,\\
u_i^\phi&=&-g^{t\phi} + g^{\phi\phi} \frac{L_i}{m} \ ,\\\nonumber
u_i^\theta&=&0 \ .
\end{eqnarray}
The radial part of the motion can be easily determined from the relation $g_{\alpha\beta} \  u_i^\alpha \ u_i^\beta=-1$ with $i=1,2$ and have the following explicit form 

\begin{eqnarray}
u^r_1=\sqrt{\frac{f}{H} \left[L^2 \left(-\left(\frac{H}{r^2}-\frac{\omega^2}{f
			H}\right)\right)-\frac{2 L \omega}{f H}+\frac{1}{f H}-1\right]}\, .
\end{eqnarray}

\textcolor{black}{For the study of particle collision one needs to calculate so-called center-of-mass energy $E_{CM}$ of the colliding particles which defines the energy of particles measured in the CM reference frame.
The CM energy of two neutral particles is defined by $E^2_{CM}=-g_{\alpha\beta} \  \tilde{p}^{\alpha}_{\rm tot} \  \tilde{p}^{\beta}_{\rm tot}$, which reduces to its classical form $E_{CM}=E_1+E_2$ in the case when the spacetime is flat and the velocity of particles is much smaller as compared to the speed of light in vacuum.} The general form of CM energy is defined by 
\begin{eqnarray} \label{CME}
\nonumber
E^2_{CM}&=&-m^2 g_{\alpha\beta} (u_1^\alpha+u_2^\alpha) (u_1^\beta+u_2^\beta)\\
&=& 2 m^2 (1-g_{\alpha\beta} u_1^\alpha u_2^\beta) \, ,
\end{eqnarray}
or explicitly
%
%\begin{widetext}
	\begin{eqnarray}
	\frac{E^2_{CM}}{2 m^2}&=&1-\frac{\left({(\mathcal{L} \omega-1)^2-\frac{f H \left(H \mathcal{L}^2+r^2\right)}{r^2}}\right)^{1/2}
		}{f
		\left(\frac{(\mathcal{L} r \omega+r)^2-f H \left(H \mathcal{L}^2+r^2\right)}{H^2 r^2}\right)^{-1/2} }\nonumber\\&&+ \frac{L\mathcal{L}^2 \omega^2-1}{fH}+\frac{H \mathcal{L}^2}{r^2}\, ,
	\end{eqnarray}
%\end{widetext}
%
for the spacetime considered here.

\begin{figure*}[t!]
	\begin{center}
		\includegraphics[scale=0.6]{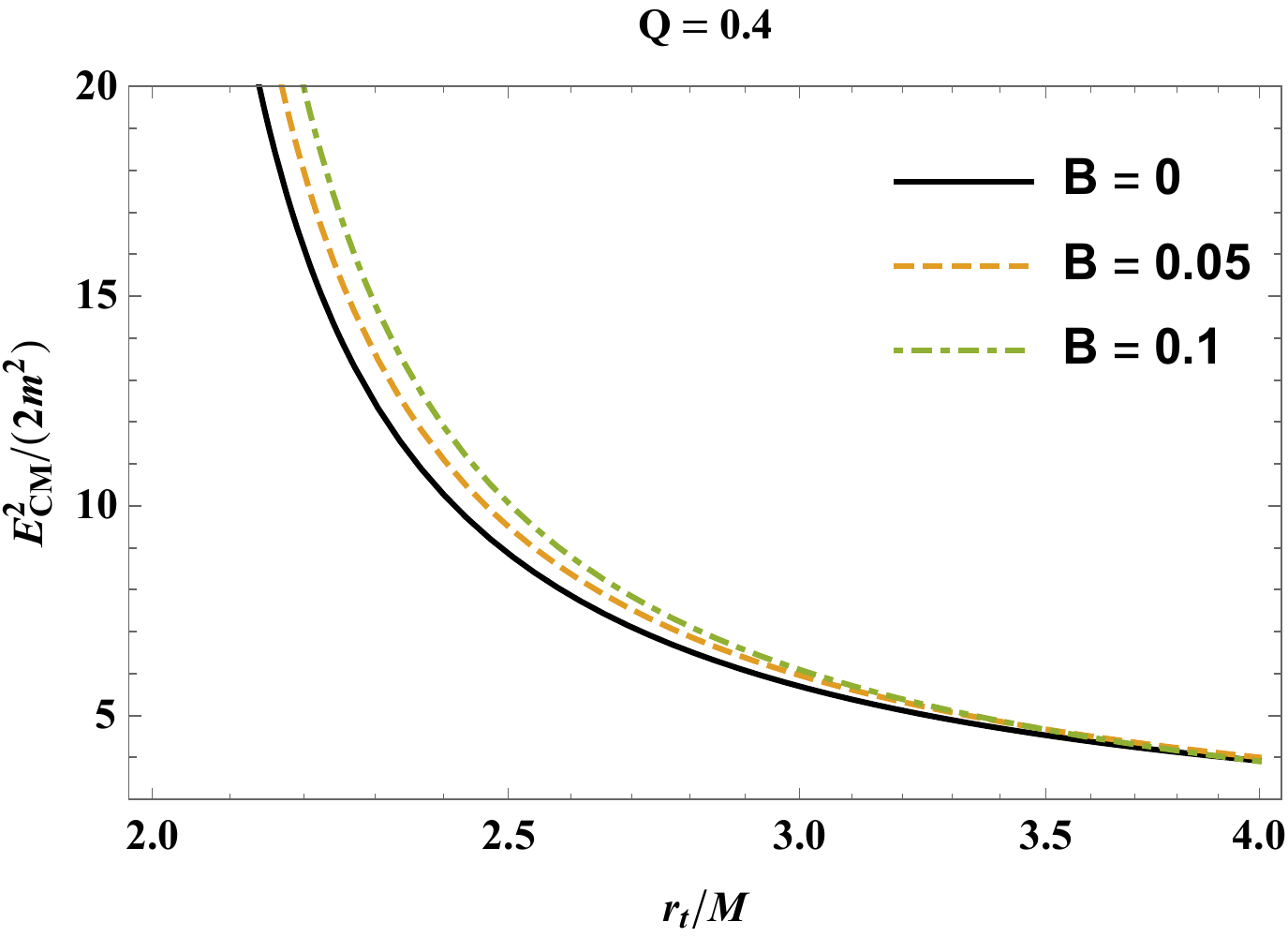}
		\includegraphics[scale=0.6]{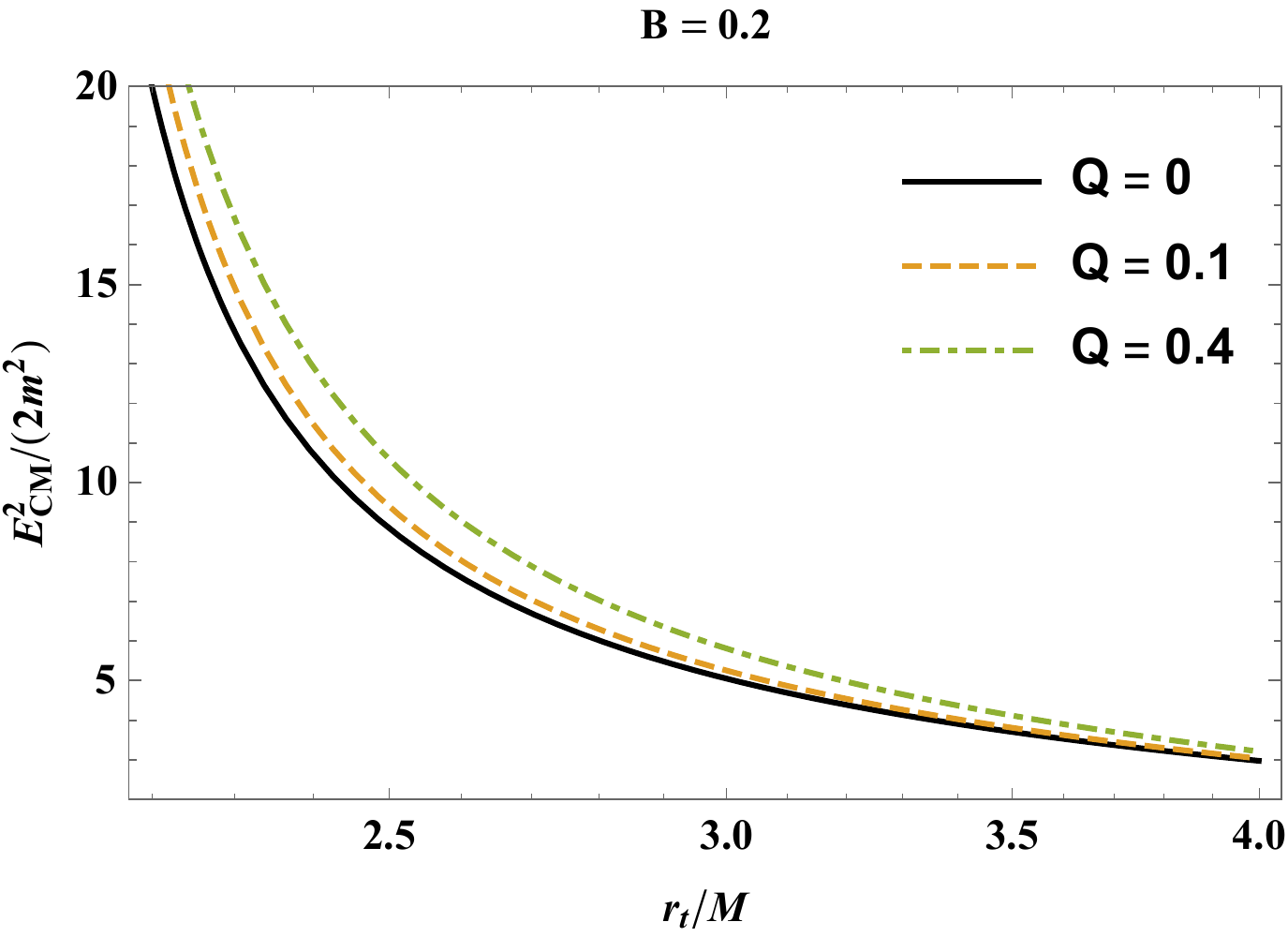}
	\end{center}
	\caption{The radial dependence of the CM energy $E_{CM}$ of the colliding neutral particles for the fixed $Q$ and various values of $b$ (left panel) and the opposite (right panel).}\label{fig3}
\end{figure*}

It is interesting to study the collision process at the turning point of particles, and thus the following condition holds well:
\begin{eqnarray} \label{tpoint}
u^r(r_t,{\cal L})=0 \, .
\end{eqnarray}
The condition in the above leads to the following form for angular momentum of test particle 
	\begin{eqnarray}
	{\cal L}(r_t)=\frac{\sqrt{-f^2 H^3 r_t^2+f H^2 r_t^2+f H w^2 r_t^4}-w r_t^2}{f H^2-w^2 r_t^2}\, ,
	\end{eqnarray}
where $r_t$ defines the radial coordinate of turning point of the test particle. The center of mass energy $E_{CM}$ of two colliding test particles as a function of $r_t$ is presented in Fig.~\ref{fig3}. As can be seen from Fig.~\ref{fig3} that $E_{CM}$ goes up exponentially as $r_t$ approaches the black hole horizon. Thus, one can see that $E_{CM}$ becomes higher for large values of magnetic field and electric charge of the black hole.

\subsection{ Collision of charged particles on circular orbits with radially falling neutral ones }

In this part, we consider that one of two colliding particles is neutral particle falling from infinity and the other is electrically charged particle moving in the equatorial plane of the magnetized Reissner-Nordstr\"{o}m black hole. As always, we assume that both test particles move in the equatorial plane and have the same masses $m_1=m_2=m$.  For the freely falling neutral particle, one can write
\begin{eqnarray}\nonumber
u_2^t&=&-g^{tt} \ ,\\
u_2^r&=&\sqrt{g^{rr} (-1-g^{tt})} \ ,\\\nonumber
u_2^\theta&=&u_2^\phi=0 \, .
\end{eqnarray}
For the charged particle making circular revolutions around the black hole, the four-velocity reads as
\begin{eqnarray}\nonumber
u_1^t&=&-g^{tt} \mathcal{E}(r,b,Q) \ ,\\
u_1^r&=&u_1^\theta=0 \ ,\\\nonumber
u_1^\phi&=&g^{\phi\phi} \mathcal{L}(r,b,Q) \, ,
\end{eqnarray}
with $\mathcal{E}(r,b,Q)$ and $\mathcal{L}(r,b,Q)$ being the energy and the angular momentum of the charged particle determined by $V_{eff}(r)=0$ and $V_{eff}'(r)=0$. Note here that we have defined $b=qBM/m$.

\begin{figure*}[t!]
	\begin{center}
		\includegraphics[scale=0.6]{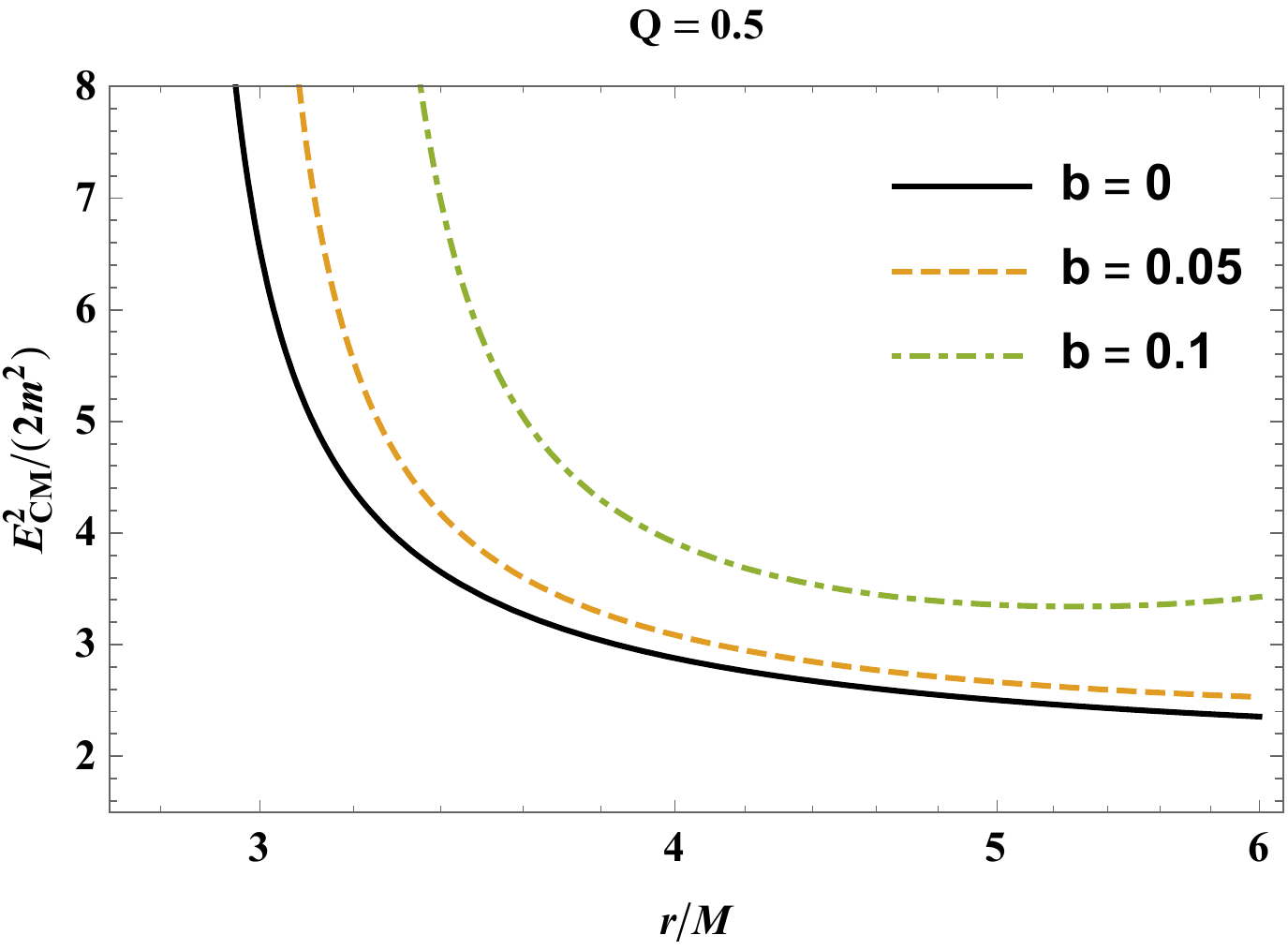}
		\includegraphics[scale=0.62]{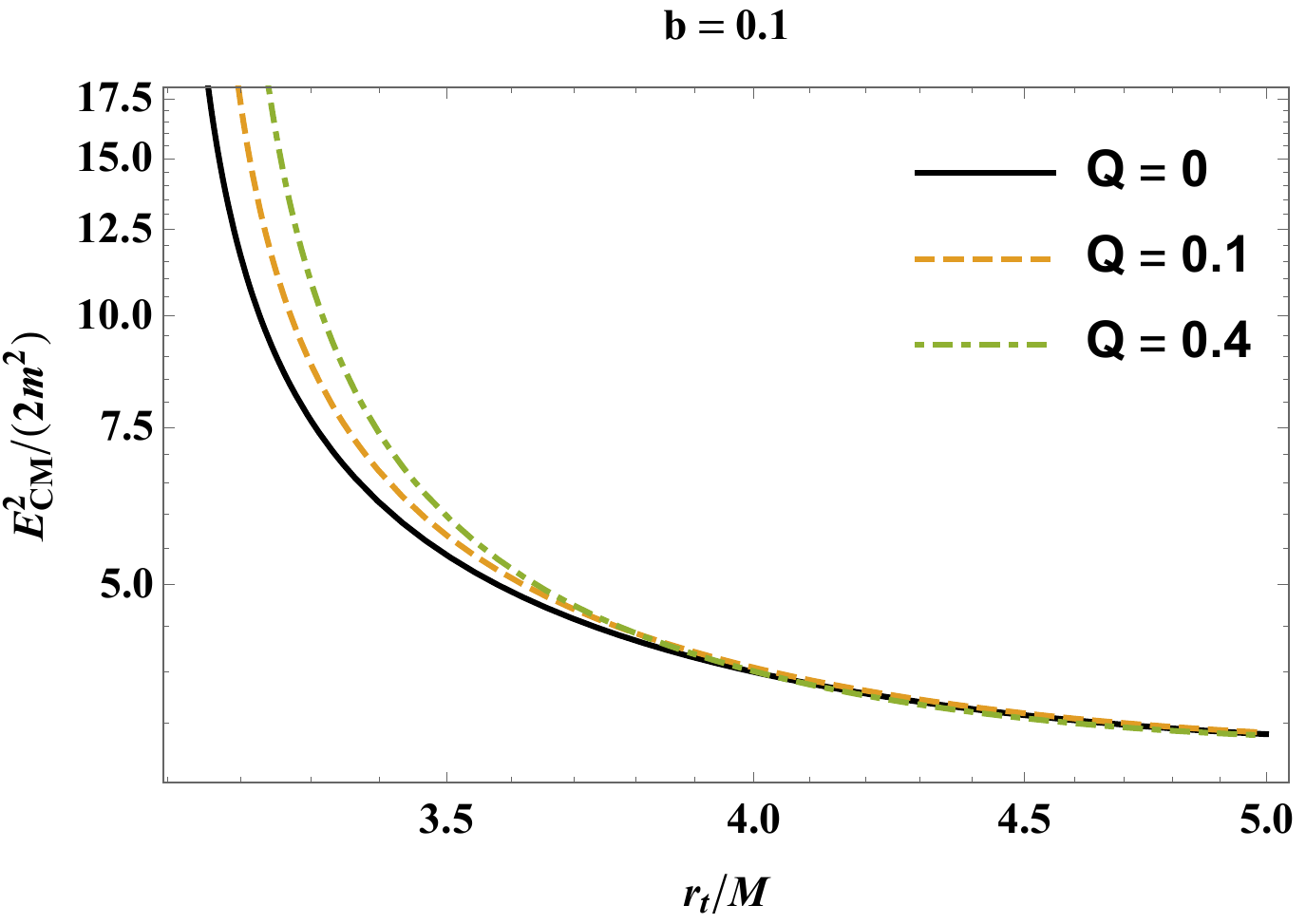}
	\end{center}
	\caption{Dependence of the CM energy $E_{CM}$ from the radius of collision of two test particles for the different values of magnetic parameter $b$ and electric charge $Q$ of the magnetized Reissner-Nordstr\"{o}m black hole.}\label{fig4}
\end{figure*}

The dependence of the CM energy of these two colliding particles from the radius of circular orbit is shown in Fig.~\ref{fig4} as stated by Eq.~ (\ref{CME}). From Fig.~\ref{fig4}, it is clearly seen that the effect of magnetic field around the black hole is stronger as compared to the electric charge of black hole, regardless of the fact that they both increase the CM energy of the colliding test particles.
\\

\section{Conclusions}
\label{Sec:conclusion}
We studied the dynamics of neutral and charged test particles around the magnetized Reissner-Nordstr\"{o}m black hole. Taking into account electromagnetic field of magnetized black hole in its vicinity we showed that the combined effects of black hole electric charge and magnetic field drastically affect the ISCO radius, thus shrinking its values. We also showed that the value of the ISCO radius depends on the sign of (positively and negatively) charged particle, i.e.  the ISCO initially increases (decreases) and then radically alters with increasing both black hole electric charge and test particle charge. This happens because the repulsive (attractive) Coulomb force dominates over the Lorentz force as that of black hole magnetic field. In fact, regardless of much progress in the astronomical observations distinguishing departures from any two black hole geometries is still impossible for distant observer by using astronomical observations allowing to analyze electromagnet radiations emitted by accretion orbiting around black hole. Thus, the study of the degeneracy for the location of the ISCO radius between the Kerr and the magnetized Reissner-Nordstr\"{o}m black hole geometries has value as it helps to understand the qualitative aspects of the magnetized Reissner-Nordstr\"{o}m black hole geometry. We found that the combined effects of black hole electric charge and magnetic field can mimic black hole spin up to $a/M\approx 0.9$.  This would of course be a theoretical model it does however play a decisive role to explain the geodesics around the magnetized Reissner-Nordstr\"{o}m black hole.

From the investigation of the CM energy extracted by particle collisions, we found that an increase in the value of both parameters of the magnetized black hole causes to the increase of the CM energy of the colliding two particles in specific collision scenarios considered here. The point to be noted here is that in the near regions close to the black hole the CM energy of colliding particles goes up exponentially, thus providing the difference from the one in Schwarzschild case and allowing to distinguish the magnetized Reissner-Nordstr\"{o}m black hole from the latter one.

\section*{Acknowledgments}
S.S. acknowledges the support of Uzbekistan Ministry for Innovative Development. B.N. acknowledges support from the China Scholarship Council (CSC), grant No. 2018DFH009013. AA is supported by PIFI fund of Chinese Academy of Sciences. 

\appendix

\bibliographystyle{apsrev4-1}  %% BibTeX style
\bibliography{gravreferences,gravreferences1}

\end{document}